\documentclass[floatfix, twocolumn, prb]{revtex4-2}

\usepackage{multirow}
\usepackage{epsfig}
\usepackage{graphicx}
\usepackage{amsmath}
\usepackage{amssymb}
\usepackage{bm}
\usepackage{dcolumn}
\usepackage{braket}
\usepackage{bbold}
\usepackage{ulem}
\usepackage{soul}
\usepackage{rotating}
\usepackage{tikz,pgfplots,gensymb}
\usepackage[hidelinks]{hyperref}
\usepackage{xcolor}
\usepackage{pifont}
\usepackage{array}
\usepackage{natbib}
\usepackage{longtable}
\usepackage{color, colortbl}
\usepackage{float}

\definecolor{Gray}{gray}{0.9}
\definecolor{LightCyan}{rgb}{0.88,1,1}
\definecolor{Blue}{rgb}{0,0,1}

\pgfplotsset{compat=1.8}
\usetikzlibrary{pgfplots.groupplots}

\hypersetup{
    colorlinks,
    linkcolor={red!50!black},
    citecolor={blue!50!black},
    urlcolor={blue!80!black}
}

\begin{document}

\title{Tunable anomalous Hall and Nernst effects in MM'X compounds}

\author{Ilias Samathrakis}
\email[corresp.\ author: ]{iliassam@tmm.tu-darmstadt.de}
\author{Nuno Fortunato}
\author{Harish K. Singh}
\author{Chen Shen}
\author{Hongbin Zhang}

\affiliation{Institute of Materials Science, TU Darmstadt, 64287 Darmstadt, Germany}

\begin{abstract}
Based on first-principles calculations, the anomalous Hall (AHC) and anomalous Nernst conductivities (ANC) of the XMnP (X=Ti, Zr, Hf) compounds were evaluated, and the possibility to tailor such properties by changing the magnetization directions was also investigated. We observed that nodal lines and small gap areas consist the dominant contributions to AHC and ANC, where the significant variations in AHC and ANC when altering the magnetization direction originate from the whole Brillouin zone. Our studies gave a promising clue on engineering magnetic intermetallic compounds for tunable transverse thermoelectric applications.
\end{abstract}

\maketitle

\section{\label{INTRO}Introduction}

\par The anomalous Hall effect (AHE) describes the generation of a transversal current, perpendicular to the electric field, in the absence of an external magnetic field~\cite{vzutic2004spintronics,nagaosa2010anomalous}. Its thermal counterpart, dubbed anomalous Nernst effect (ANE)~\cite{behnia2016nernst,lee2004anomalous}, is achieved if the electric field is replaced by a thermal gradient. Ferromagnetic materials were the first observed to possess non zero AHE and ANE. It was later found that antiferromagnetic materials with noncollinear structure can exhibit large anomalous Hall (AHC) and Nernst conductivities (ANC) with the most representative examples being the Mn$_3$X with X=(Ge,Ga,Sn,Rh,In)~\cite{chen2014anomalous,zhang2018spin,nayak2016large,ganguly2011augmented,
yang2017topological,kubler2014non,li2017anomalous,guo2017large,kiyohara2016giant,ikhlas2017large,vzelezny2017spin} and the Mn$_3$XN with X=(Ga,Zn,Ag,Ni)~\cite{zhou2020giant,gurung2019anomalous,huyen2019topology} families. Very recently, it was proposed that the collinear antiferromagnet RuO$_2$ can induce nonzero AHC~\cite{feng2020observation}. The existence of such topological properties in these magnetic materials can be traced back to the appearance of weyl nodes and nodal lines to the electronic structure~\cite{yang2017topological}, being non-accidental linear touching points (Weyl semimetals) or lines (nodal semimetals) close to the Fermi energy, that cause non zero Berry curvature, acting as a fictitious magnetic field, leading to non vanishing effects. From the practical point of view, compounds with large values of AHC and ANC are promising candidates for data storage~\cite{hasegawa2015material} and data transfer~\cite{huang2011intrinsic} devices that can lead to further spintronics applications.

\par The MM'X (X=main group element, M and M'=metal) are a widespread class of compounds that can form in orthorhombic TiNiSi, along with the less common Ni$_2$In and LiGaGe hexagonal structures. They accumulate several intriguing properties such as the martensitic transition~\cite{johnson1975diffusionless,szytula1981crystal,jeitschko1975high} and have attracted scientific attention due to interesting properties such as magnetoresistance~\cite{yu2006large,barandiaran2009effect}, magneto-strain effect~\cite{ullakko1996large,wu1999giant,kainuma2006magnetic} and magnetocaloric effect~\cite{krenke2005inverse,gutfleisch2011magnetic,tegus2002transition,liu2012giant}, making them promising candidates for solid-state magnetic cooling applications ~\cite{liu2013giant,glanz1998making,gschneidnerjr2005recent}. Besides, the members of the MM'X family have been reported to host large anomalous Hall conductivity (AHC) and anomalous Nernst conductivity (ANC), such as the primordial research on their topological properties of ZrMnP and HfMnP compounds, showing large AHC values of the order of 1000 $S/cm$ and 1500 $S/cm$ respectively~\cite{singh2021anisotropic}, due to the presence of nodal lines, rendering them as promising materials~\cite{lamichhane2016discovery}.

\par In this work, we applied first principles calculations to evaluate the AHC and ANC of XMnP (X=Ti,Zr,Hf) compounds and altered the magnetization direction to tune such transport properties. Our results demonstrated that both AHC and ANC originated from the combination of nodal lines and small gap areas present in the whole Brillouin zone. It was further observed that the magnetization direction significantly affected the evaluated AHC and ANC with their differences being distributed throughout the whole Brillouin zone

\section{Numerical details}

\par Our first principle calculations were performed using the projector augmented wave method, as implemented in VASP software ~\cite{Kresse:1993}. The exchange-correlation functional was approximated using the generalized gradient approximation (GGA) as parameterized by Perdew-Burke-Ernzerhof~\cite{Perdew:1996}. A $\Gamma$-centered kmesh sampling of density 50 in respect to all lattice dimensions as well as an energy cutoff of 500eV were used. Spin orbit coupling (SOC) was included in all calculations. The DFT-Bloch wave functions were projected on maximally localized wannier functions (MLWF), using the wannier90 software~\cite{Mostofi:2008}. The MLWFs were constructed for $spd$-orbitals of Mn, Ti and Hf, $sp$-orbitals of  $spd$-orbitals of Zr, 
generated on a kmesh of 40 and the frozen window was computed based on the partial density of states, according to the methodology described in~\cite{Zeying:2018}. 
The AHC was evaluated using Wanniertools~\cite{wanniertools:2018} by integrating the Berry curvature according to the formula:
\begin{align}
& \sigma_{ij} = \frac{e^2}{\hbar} \int \frac{d\mathbf{k}}{\left(2\pi \right)^3} \sum_n f\big[ \epsilon\left(\mathbf{k}\right) - \mu \big] \Omega_{n,ij} \left(\mathbf{k}\right), \label{aheeq} \\
& \Omega_{n,ij}\left(\mathbf{k}\right) = -2 Im \sum_{m \neq n} \frac{\bra{\mathbf{k}n}u_i\ket{\mathbf{k}m}\bra{\mathbf{k}m}u_j\ket{\mathbf{k}n}}{\big[\epsilon_m\left(\mathbf{k}\right)-\epsilon_n\left(\mathbf{k}\right)\big]^2}
\end{align}
where $f$ is the Fermi distribution function, $\mu$ the Fermi energy, n and m the occupied and empty Bloch band with $\epsilon_n\left(\mathbf{k}\right)$ and $\epsilon_m\left(\mathbf{k}\right)$ being their energy eigenvalues and $v_i$ the velocity operator. The ANC was computed using an in-house developed python script based on the formula~\cite{Xiao:2010}:
\begin{equation}
\alpha_{ij} = -\frac{1}{e} \int d\epsilon \frac{\partial f}{\partial \mu} \sigma_{ij}\left(\epsilon\right) \frac{\epsilon-\mu}{T} 
\label{aneeq}
\end{equation}
where $e$ is the charge of the electron, $T$ the temperature and $\epsilon$ the energy point within an energy grid of 1000 points at a range $\big[-0.5,0.5\big]$ eV in respect to the Fermi energy.

\section{\label{RES}Results and Discussion}
\par
\begin{figure}
\includegraphics[width=0.4\textwidth]{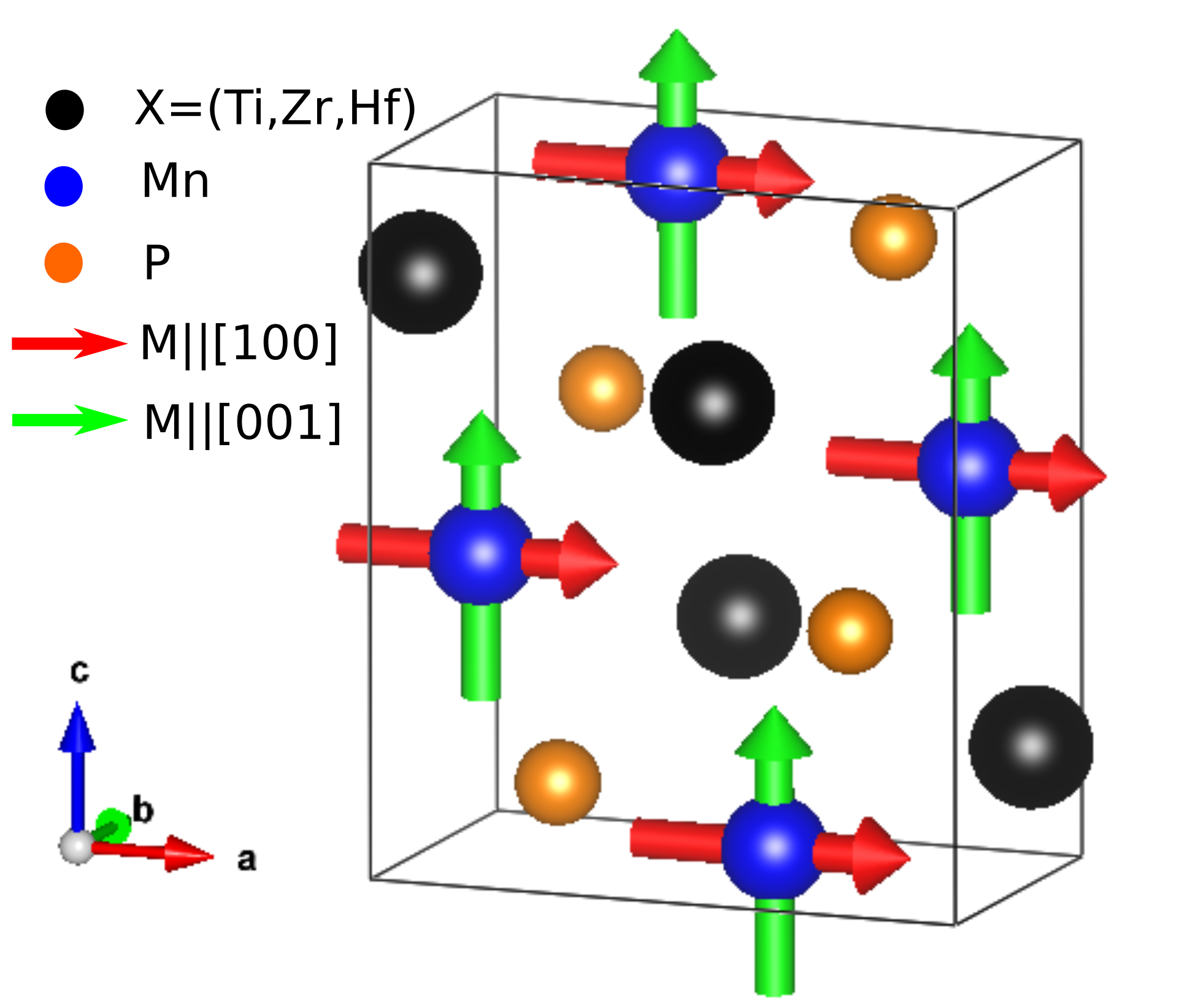}
\caption{Crystal structure of XMnP (X=Ti,Zr,Hf) with magnetic moments of Mn atoms pointing along [100]-direction (red) and along [001]-direction (green) }
\label{crystal_str}
\end{figure}

 The transition metal pnictides under consideration crystallize in the orthorhombic $Pnma$ (No. 62) space group with lattice parameters being equal to $a=6.43\AA$, $b=3.63\AA$, $c=7.51\AA$ for ZrMnP and $a=6.37\AA$, $b=3.60\AA$, $c=7.45\AA$ for HfMnP, obtained from X-rays experiments in Ref.~\cite{singh2021anisotropic} and Ref.~\cite{lamichhane2016discovery} and $a=6.15\AA$, $b=3.47\AA$, $c=7.20\AA$ for TiMnP, available to the inorganic crystal structure database (ICSD). The magnetic moments originate exclusively from Mn atoms and have been calculated to be $1.74\mu_B$, $2.06\mu_B$ and $1.95\mu_B$ for TiMnP, ZrMnP and HfMnP respectively, in reasonable agreement with the values of $1.8\mu_B$ and $2.0\mu_B$ at $T=2K$ for ZrMnP and HfMnP respectively, calculated in Ref.~\cite{singh2021anisotropic}. The input values of the lattice constants and the magnetic moments of Mn atoms are summarized in the Tab.~\ref{data}. Our DFT calcualtions were performed on two magnetic configurations, namely with magnetic moments aligned along [100]- and along [001]-axis, as illustrated in Fig.~\ref{crystal_str}. These two discrepant energetically distinguishable magnetic configurations exhibit a magneticrystalline anisotropy energy (MAE) of 0.165 meV per Mn atom and 0.43 meV per Mn atom in favor of [100]-direction for ZrMnP and HfMnP respectively. Our results are in good agreement with the values of 0.136 meV per Mn atom and 0.47 meV per Mn atom reported in Ref.~\cite{lamichhane2016discovery} for ZrMnP and HfMnP respectively, where altering the magnetic configuration is possible by means of external field. In addition TiMnP exhibits a MAE of 0.05 meV per Mn atom in favor of [100]-direction.

\begin{table*}
\caption{Summary of lattice constants used as input to our calculations and calculated output of magnetization, AHC and ANC compared to the existing literature for TiMnP, ZrMnP, HfMnP.}
\begin{tabular}{|c|c|c|c|c|c|c|c|c|c|c|}
\hline
\multicolumn{1}{|c|}{Compound} & \multicolumn{3}{|c|}{Lattice constant $(\AA)$} & \multicolumn{2}{|c|}{Magnetization ($\mu_B$)} & \multicolumn{3}{|c|}{AHC ($S/cm$)} & \multicolumn{2}{|c|}{ANC ($A/mK$)} \\
\hline
 & a & b & c & calculated & theory & $\sigma_x,M||[100]$ & $\sigma_x$ theory & $\sigma_z,M||[001]$ & $\alpha_x,M||[100]$ & $\alpha_z,M||[001]$ \\
\hline
TiMnP & 6.15 & 3.47 & 7.20 & 1.74 & - & 624 & - & 224 & -2.08 & 0.02 \\
ZrMnP & 6.43 & 3.63 & 7.51 & 2.06 & 1.80~\cite{singh2021anisotropic} & 962 & 1000~\cite{singh2021anisotropic} & 778 & 0.21 & -3.43 \\
HfMnP & 6.37 & 3.60 & 7.45 & 1.95 & 2.00~\cite{singh2021anisotropic} & 1516 & 1500~\cite{singh2021anisotropic} & 769 & 0.01 & -2.97 \\
\hline
\end{tabular}
\label{data}
\end{table*}

\par
\begin{figure}
\includegraphics[width=0.5\textwidth]{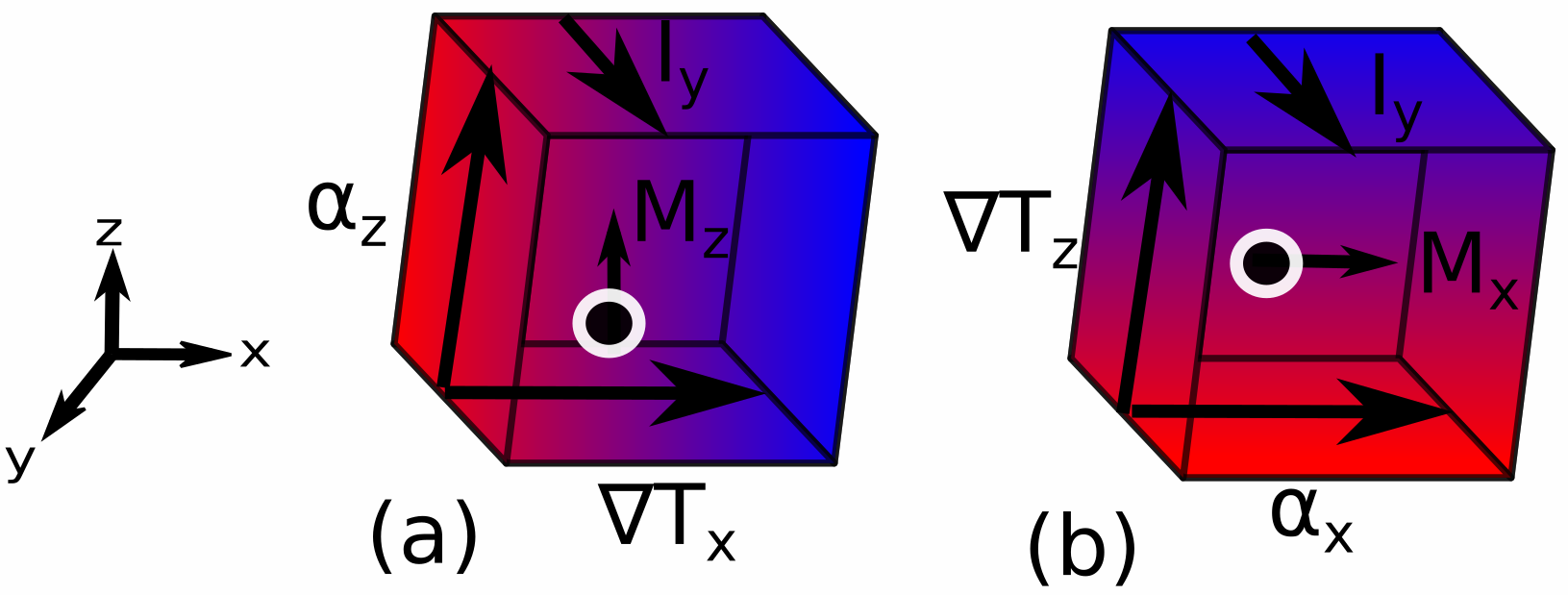}
\caption{Schematic illustration of the induced anomalous Nernst conductivity direction change for (a) magnetization direction parallel to [001] axis (b) magnetization direction parallel to [100] axis.}
\label{fig_ND}
\end{figure}

\begin{figure*}
\includegraphics[width=0.85\textwidth]{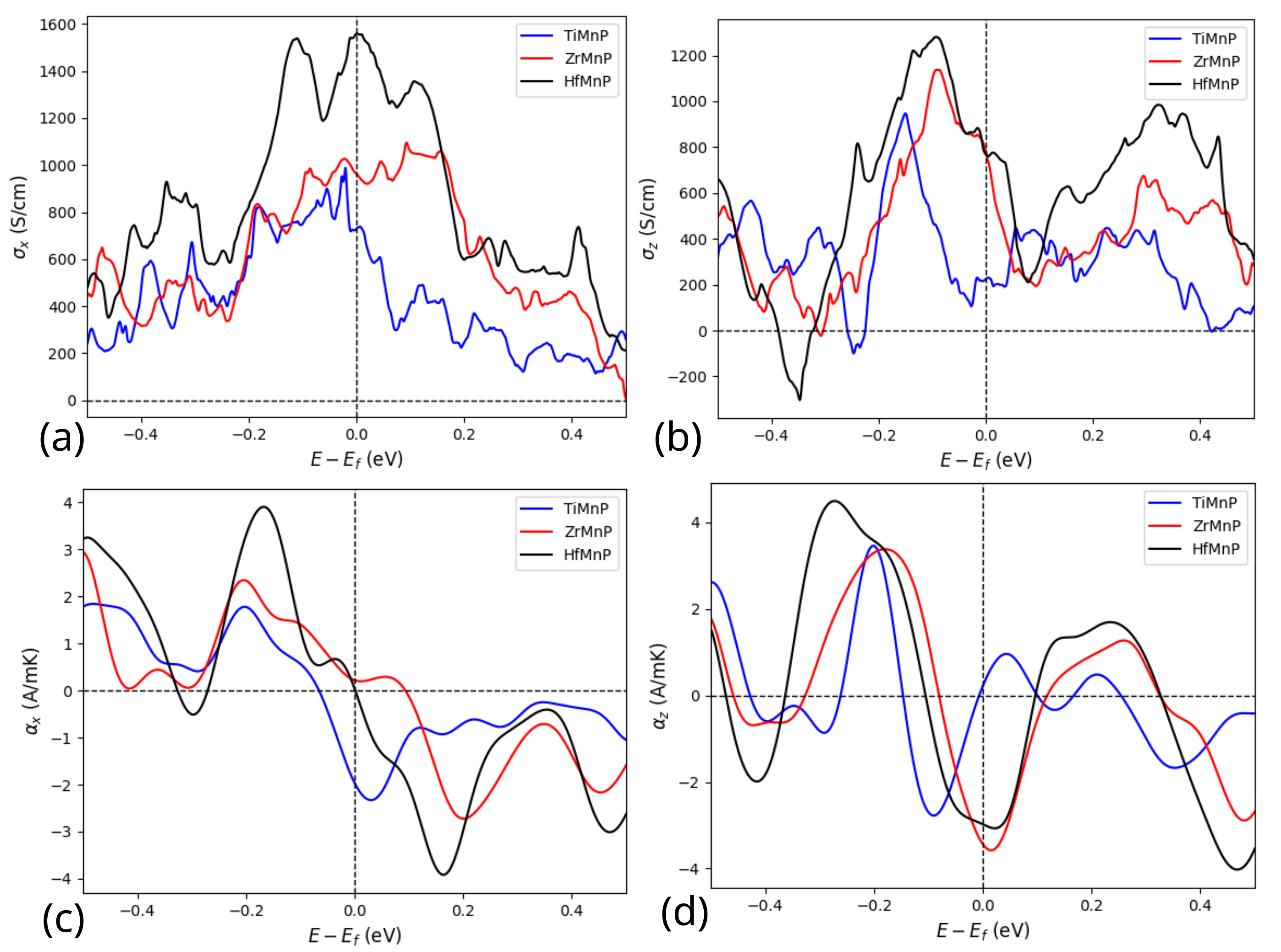}
\caption{(a) $\sigma_x$ for XMnP (X=Ti,Zr,Hf) with magnetization direction parallel to [100]-axis. (b) $\sigma_z$ for XMnP (X=Ti,Zr,Hf) with magnetization direction parallel to [001]-axis. (c) $\alpha_x$ for XMnP (X=Ti,Zr,Hf) with magnetization direction parallel to [100]-axis. (d) $\alpha_z$ for XMnP (X=Ti,Zr,Hf) with magnetization direction parallel to [001]-axis.}
\label{fig1}
\end{figure*}

 Significant changes in AHC and ANC values are observed by altering the magnetization direction of the system. A schematic illustration of a thermopile is shown in Fig.~\ref{fig_ND}. The direction of the induced ANC is depicted for fixed directions of thermal gradient, magnetization and electric current. The same principles apply to AHC with the electric field replacing the thermal gradient, demonstrating the geometry in which the induced AHC/ANC is parallel to the magnetization. An interesting question is weather the AHC  and ANC vectors of the XMnP (X=Ti,Zr,Hf) compounds of the MM'X family are modified by altering the magnetization direction. In order to address, we calculated AHC and ANC for the magnetization direction being along [100] and [001], as illustrated in Fig.~\ref{crystal_str}. An increase of AHC with changing the magnetization direction from [001] to [100] is observed in all compounds. Namely, for TiMnP there is an almost 3 times increase (from 224 $S/cm$ with $M||[001]$ to 624 $S/cm$ with $M||[100]$). More interestingly, the calculated value of 778 $S/cm$ with $M||[001]$ jumps to 962 $S/cm$ with $M||[100]$ in ZrMnP and the 769 $S/cm$ with $M||[001]$ skyrockets to 1516 $S/cm$ with $M||[100]$ for HfMnP, being in excellent agreement with the reported theoretical (experimental) values of 1000 $S/cm$ (900 $S/cm$) and 1500 $S/cm$ (1400 $S/cm$) for ZrMnP and HfMnP respectively~\cite{singh2021anisotropic}, calculated for $M||[100]$, as shown in Fig.~\ref{fig1}(b) and Tab.~\ref{data}. It is further noted that the calculated value of AHC of HfMnP is larger than many compounds discussed in Ref.~\cite{samathrakis2021enhanced} including the reported value of -1310 $S/cm$ for Co$_3$Sn$_2$S$_2$ in Ref.~\cite{wang2018large} and many Heusler compounds reported in Ref.~\cite{noky2020giant}. Surprisingly, the calculated ANC exhibits a peculiar behavior, where for TiMnP the finite value of -2.08 $A/mK$ for the [100] direction becomes 0.02 $A/mK$ for the [001] direction. The opposite behavior is observed in HfMnP and ZrMnP where the small values of 0.21 $A/mK$ and 0.01 $A/mK$ respectively increase (in absolute value) to -2.97 $A/mK$ and -3.43 $A/mK$, which is among the largest reported in Ref.~\cite{wang2018large}. Similar ANC changes have been studied in non-collinear antiperovkites in Ref.~\cite{singh2022giant}
These discrepant AHC/ANC values serve an interesting playground for AHC/ANC manipulations by altering the magnetization direction of the compounds, as illustrated in Fig.~\ref{fig_ND}, that can be useful for transverse thermoelectric generation having great potential for energy harvesting and heat sensing applications~\cite{zhou2021seebeck,uchida2021transverse,sakai2020iron,holanda2021thermal}.

\par It is observed that the symmetry of the Berry curvature plays the predominant role in determining the numerical values of the AHC and ANC tensor elements, given by Eq.~\ref{aheeq} and Eq.~\ref{aneeq} respectively. Since the Berry curvature behaves as a pseudovector under symmetry operations, it is possible to predetermine its value based on the magnetic point group of the compound, according to the formula~\cite{seemann2015symmetry}:
\begin{equation}
s\mathbf{\Omega}\left(\mathbf{k}\right) = \pm det\left(\mathbf{D}\left(R\right)\right) \mathbf{D}\left(R\right) \mathbf{\Omega}\left(s^{-1}\mathbf{k}\right),
\label{bcsym}
\end{equation}
where $\mathbf{\Omega}\left(r\right)$ denotes the pseudovector Berry curvature, $\mathbf{D}\left(R\right)$ the three-dimensional matrix representation of a symmetry operator without the translation part and finally $s$ the matrix representation of an arbitrary symmetry operation. The underlying compounds of XMnP (X=Ti,Zr,Hf) with the magnetization direction parallel to [100]-axis, belong to the magnetic space group $Pnm'a'$ (62.447), in which the presence of $2_{100}$ symmetry operation, after summing over all Brillouin zone, forces $\Omega_y=\Omega_z=0$, following the relations:
\begin{align}
& \Omega_x\left(k_x,-k_y,-k_z\right) = \Omega_x\left(k_x,k_y,k_z\right), \notag \\
& \Omega_y\left(k_x,-k_y,-k_z\right) = -\Omega_y\left(k_x,k_y,k_z\right), \\
& \Omega_z\left(k_x,-k_y,-k_z\right) = -\Omega_x\left(k_x,k_y,k_z\right) \notag.
\end{align}
On the other hand there is no condition implying vanishing value for $\Omega_x$, meaning that $\sigma_x$ is allowed to have finite value, as it happens. By changing the magnetization direction parallel to [001]-axis, the resulting magnetic space group alters to $Pn'm'a$ (62.446), in which the presence of $2_{001}$ leads to $\Omega_x=\Omega_y=0$ and leaving $\Omega_z$ unrestricted, after summing over all Brillouin zone, according to:
\begin{align}
& \Omega_x\left(-k_x,-k_y,k_z\right) = -\Omega_x\left(k_x,k_y,k_z\right), \notag \\
& \Omega_y\left(-k_x,-k_y,k_z\right) = -\Omega_y\left(k_x,k_y,k_z\right), \\
& \Omega_z\left(-k_x,-k_y,k_z\right) = \Omega_z\left(k_x,k_y,k_z\right) \notag.
\end{align}
It is noted that the absence of symmetry operations forcing a specific component to vanish, does not necessarily guarantee its finite value, as it was found for Co$_2$NbAl in Ref.~\cite{samathrakis2021enhanced}.  

\par Negligible ANC values are allowed even in the presence of finite AHC values evaluated at the Fermi energy. The shape of the AHC tensor is determined by the Berry curvature, as described by Eq.~\ref{bcsym}. The same principles apply to the ANC tensor therefore one would expect the same non vanishing components in both AHC and ANC tensors. While this is in principle true, there are certain conditions in which the ANC exhibits an incidental vanishing value in the presence of a finite AHC value. The first, trivial case, appears when the AHC is a finite constant for a wide range of energies around the Fermi energy. In this case, the derivative of a constant vanishes, giving negligible ANC values, as occurs in TiMnP for magnetization parallel to [001]-axis (see blue AHC curve in Fig.~\ref{fig1}(b) and the resulting blue ANC curve in Fig.~\ref{fig1}(d)). Another, less obvious, case happens if the AHC curve as a function of energy is symmetric (or almost symmetric) for a certain range around the Fermi energy. In this case, the contribution from the range below the Fermi energy will cancel out the one from the range above the Fermi energy, giving rise to vanishing ANC value, as observed in HfMnP for magnetization parallel to [100]-axis (see black AHC curve in Fig.~\ref{fig1}(a) and the resulting black ANC curve in Fig.~\ref{fig1}(c)). 

\par
\begin{figure*}
\includegraphics[width=0.7\textwidth]{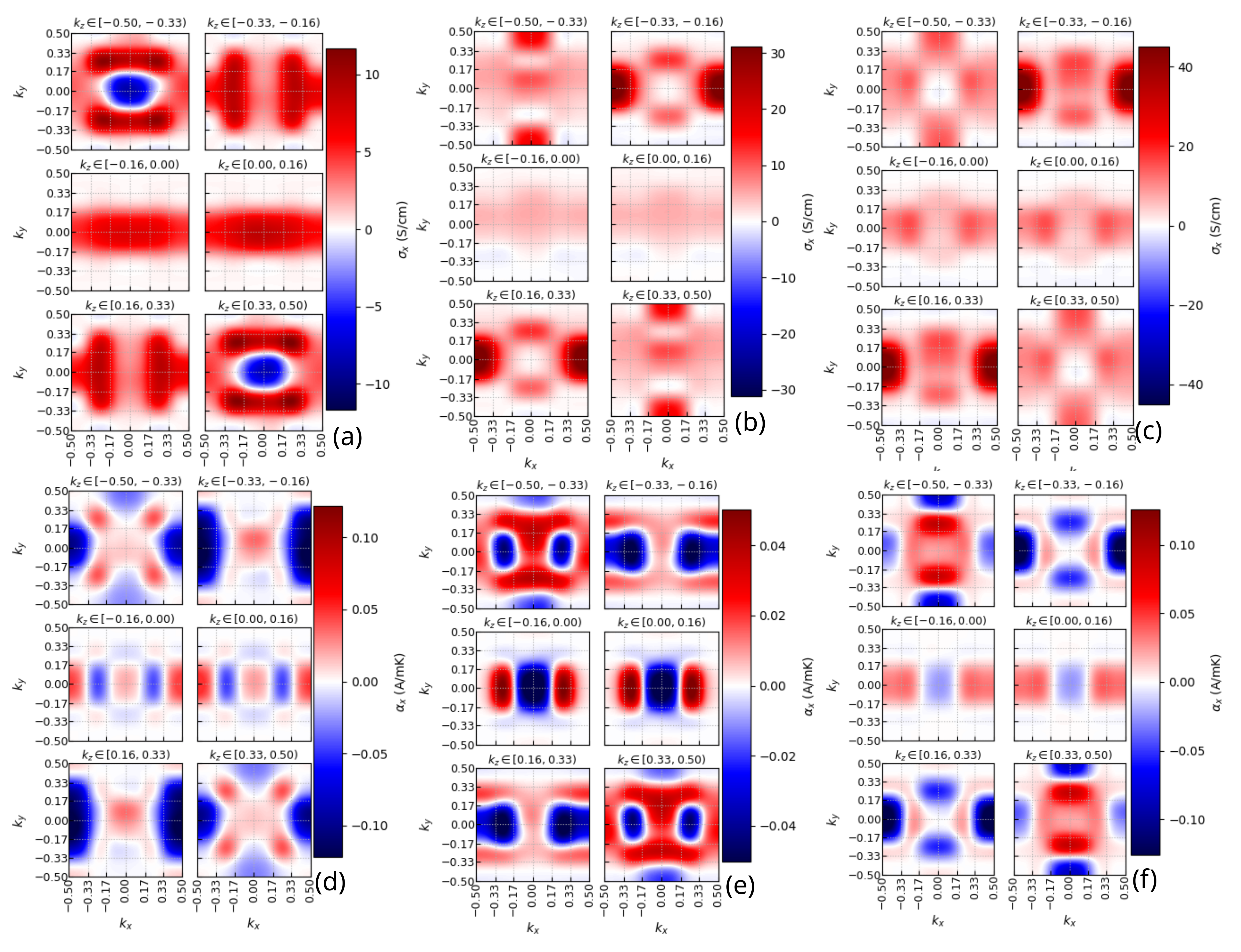}
\caption{The x-component of AHC ($\sigma_x$) and ANC ($\alpha_x$) evaluated within 216 cubes in the whole Brillouin zone for XMnP (X=Ti,Zr,Hf). (a) $\sigma_x$ for TiMnP with $M||[100]$ (b) $\sigma_x$ for ZrMnP with $M||[100]$ (c) $\sigma_x$ for HfMnP with $M||[100]$ (d) $\alpha_x$ for TiMnP with $M||[100]$ (e) $\alpha_x$ for ZrMnP with $M||[100]$ (f) $\alpha_x$ for HfMnP with $M||[100]$.}
\label{cubes}
\end{figure*}

\begin{figure*}
\includegraphics[width=0.7\textwidth]{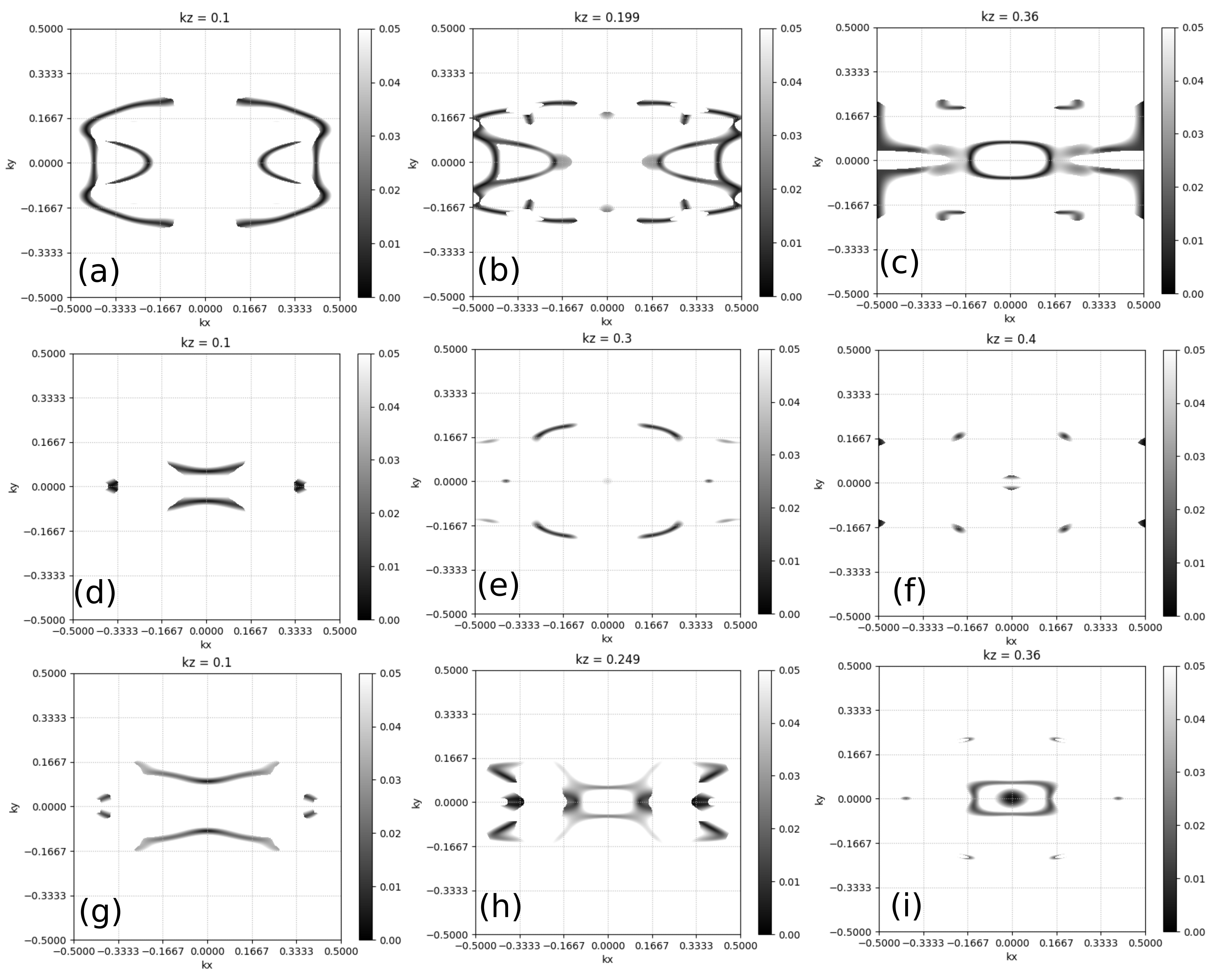}
\caption{Gap of selected $k_z$ slices of the Brillouin zone for TiMnP (a-c), ZrMnP (d-f) and HfMnP (g-i)}
\label{gapplanes}
\end{figure*}

 The AHC exhibits a disperse nature originating from the whole Brillouin zone. In order to identify the regions of dominant AHC contribution, we split the Brillouin zone of each of the underlying MM'X compounds, with magnetization direction fixed along [100]-direction, in $6 \times 6 \times 6$ cubes, within each of which the $\sigma_x$ component of the AHC tensor was calculated. The results are illustrated in Fig.~\ref{cubes}(a)-(c) for TiMnP, ZrMnP and HfMnP respectively. Even though the largest AHC contribution of ZrMnP and HfMnP is originating from the cube $k_x\in\big(0.33,0.50\big), k_y\in\big(0.00,0.16\big), k_z \in\big(0.16,0.33\big)$ (and symmetry equivalent), additional contributing parts are found throughout the whole Brillouin zone. Explicit band gap calculations reveal the main reason of contribution for each of the positive $k_z$ ranges of Fig.~\ref{cubes}(b-c), as illustrated in Fig.~\ref{gapplanes}(d-i). It is noted that the presence of several nodal lines, spread throughout the Brillouin zone is the main AHC contribution, demonstrating the results of Ref.~\cite{singh2021anisotropic}. Unlike the other two compounds, a direct comparison of Fig.~\ref{cubes}(a) with Fig.~\ref{gapplanes}(a-c) for TiMnP, reveals the presence of several nodal lines as the main origin of the enhanced AHC for the range $k_z\in\big(0.00,0.33\big)$ however, small gap areas that contribute, alongside nodal lines, are found within the range $\left(k_x,k_y\right)\in\left(0.00,0.16\right),k_z\left(0.33,0.500\right)$ of Fig.~\ref{gapplanes}(c). That is, for TiMnP the total AHC value originates from a combination of nodal lines and small gap areas. Discrepant neighboring AHC values affect the ANC that is calculated in a similar way by integrating the AHC within the specified limit of each cube, as shown in Figs~\ref{cubes}(d)-(f). Similarly, the contribution originates from a wide range of Brillouin zone. 

\par
\begin{figure*}
\includegraphics[width=0.8\textwidth]{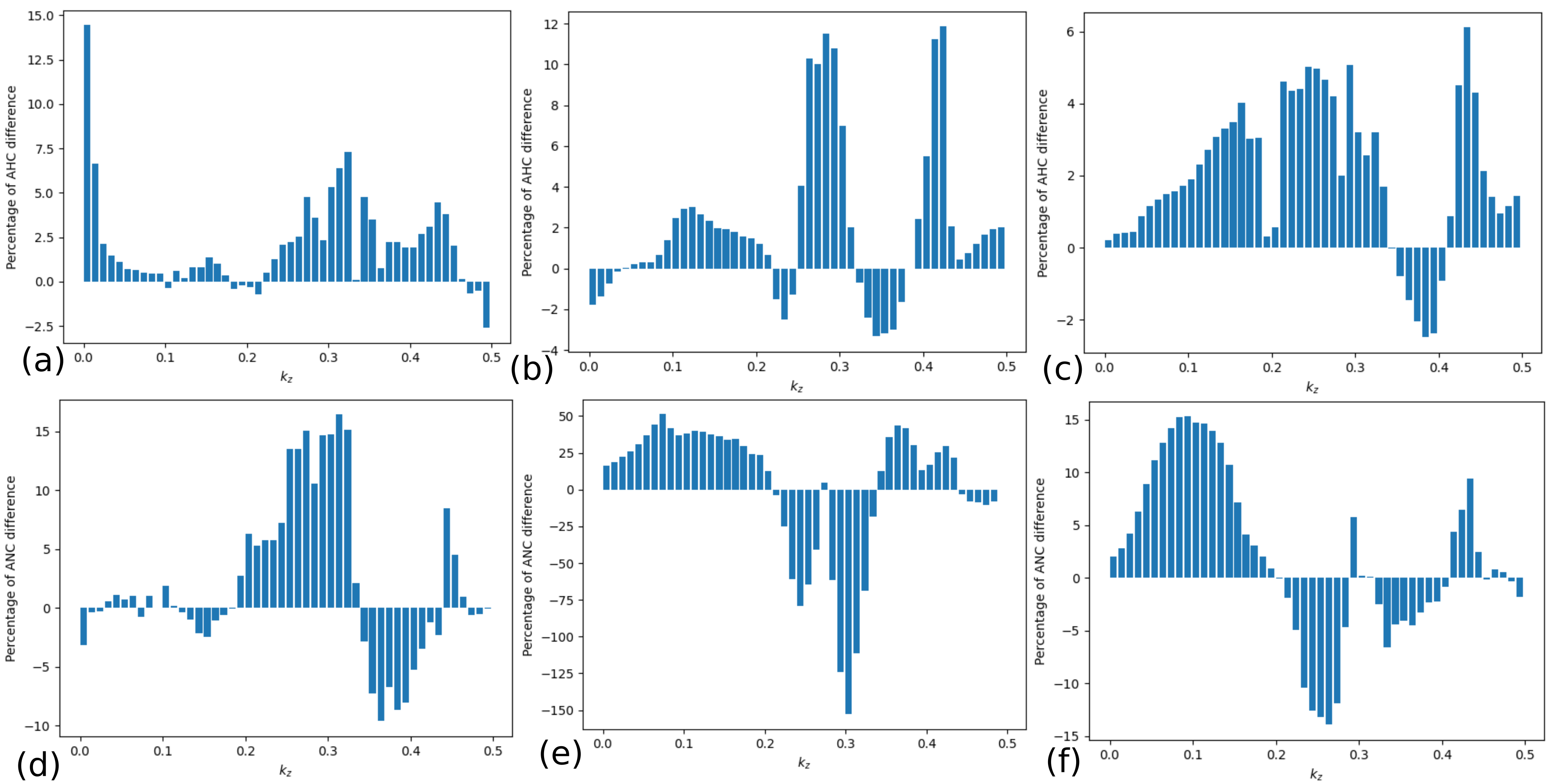}
\caption{Percentage of AHC vector change between magnetization directions [001] and [100] for $k_z \in \big[0.00,0.50\big]$ for (a) TiMnP, (b) ZrMnP, (c) HfMnP and ANC vector for (d) TiMnP (e) ZrMnP (f) HfMnP.}
\label{diff}
\end{figure*}

 Large changes of AHC \& ANC values calculated for the different magnetization directions can occur even in the absence of localized contributions. Since altering the magnetization direction affects the finite AHC component, based on the geometry of Fig.~\ref{fig_ND}, the calculated AHC changes reflect the AHC vector. Our calculations verify an increase of 441 S/cm, 182 S/cm and 693 S/cm in the AHC vector when the magnetization is altered from [001]-axis to [100]-axis for XMnP (X=Ti,Zr,Hf) respectively. Therefore, one interesting question is whether, by tuning the magnetization direction, there are emerging localized contributions originating from weyl nodes that compose the main difference in the calculated values or not. In order to identify the origin of the main difference, we split the Brillouin zone in 50 parallelepipeds for $k_z \in \big[0.0,0.50\big]$ and compute the AHC \& ANC vector change within (see Fig~\ref{diff}). It is highlighted that, for all compounds, even though there are notable differences among the contributions of several $k_z$ ranges, there is no dominant contribution, resulting in the almost uniform distribution of Fig.~\ref{diff}(a)-(f). As a result, it is demonstrated that large AHC and ANC differences are originating from small changes throughout the whole Brillouin zone instead of gigantic localized contributions.

\par What affects the AHC sign? Comparing the distribution of $\sigma_x$ component in all MM'X compounds, illustrated in Figs.~\ref{cubes}(a)-(c), we came across an interesting observation. The strong disperse negative-sign area of TiMnP within the rectangle $\left(k_x,k_y\right)\in \big[-0.166,0.166\big], k_z\in \big[0.333,0.500\big]$ (see Fig.~\ref{cubes}(a)) does not exist in any of the other compounds. In order to investigate the origin of this contribution, explicit band structure search was performed, revealing that the presence of a nodal line at the spatial vicinity of $\left(k_x,k_y\right)\approx 0.03, k_z\in \big[0.420,0.500\big] $ is responsible for this enhanced value, by contributing (including the symmetry equivalent parts) 5.6\% (35$S/cm$ out of 624$S/cm$) to the total value. The reason of the negative sign contribution though is unclear and it might be an interesting topic for further research.

\section{\label{CONC}Conclusion}

Using first principles calculations, we evaluated the AHC and ANC of XMnP (X=Ti,Zr,Hf) compounds of the MM'X family and altered the magnetization direction to tune such transport properties. The detailed symmetry analysis based on the magnetic point group of the compounds indicated the vanishing and finite values of AHC and ANC tensors, verifying the validity of our results. It was noted that nodal lines and extended gap areas formed the main AHC \& ANC contributions, demonstrating their disperse nature. Additionally, altering the magnetization direction from [001] to [100]-axis, significantly affected the AHC \& ANC values with their differences being distributed throughout the whole Brillouin zone. The observed AHC/ANC enhancements due to magnetization directions make an interesting opportunity for using the underlying MM'X members to transverse thermoelectric generation applications.

\begin{center}
\small \textbf{ACKNOWLEDGMENTS}
\end{center}

\noindent This work was financially supported by the Deutsche Forschungsgemeinschaft (DFG) via the priority programme
SPP 1666 and the calculations were conducted on the Lichtenberg high performance computer of the TU Darmstadt. We thank Dr. Ruiwen Xie for valuable discussions.

%\bibliography{bib_file}

\begin{thebibliography}{53}%
\makeatletter
\providecommand \@ifxundefined [1]{%
 \@ifx{#1\undefined}
}%
\providecommand \@ifnum [1]{%
 \ifnum #1\expandafter \@firstoftwo
 \else \expandafter \@secondoftwo
 \fi
}%
\providecommand \@ifx [1]{%
 \ifx #1\expandafter \@firstoftwo
 \else \expandafter \@secondoftwo
 \fi
}%
\providecommand \natexlab [1]{#1}%
\providecommand \enquote  [1]{``#1''}%
\providecommand \bibnamefont  [1]{#1}%
\providecommand \bibfnamefont [1]{#1}%
\providecommand \citenamefont [1]{#1}%
\providecommand \href@noop [0]{\@secondoftwo}%
\providecommand \href [0]{\begingroup \@sanitize@url \@href}%
\providecommand \@href[1]{\@@startlink{#1}\@@href}%
\providecommand \@@href[1]{\endgroup#1\@@endlink}%
\providecommand \@sanitize@url [0]{\catcode `\\12\catcode `\$12\catcode
  `\&12\catcode `\#12\catcode `\^12\catcode `\_12\catcode `\%12\relax}%
\providecommand \@@startlink[1]{}%
\providecommand \@@endlink[0]{}%
\providecommand \url  [0]{\begingroup\@sanitize@url \@url }%
\providecommand \@url [1]{\endgroup\@href {#1}{\urlprefix }}%
\providecommand \urlprefix  [0]{URL }%
\providecommand \Eprint [0]{\href }%
\providecommand \doibase [0]{https://doi.org/}%
\providecommand \selectlanguage [0]{\@gobble}%
\providecommand \bibinfo  [0]{\@secondoftwo}%
\providecommand \bibfield  [0]{\@secondoftwo}%
\providecommand \translation [1]{[#1]}%
\providecommand \BibitemOpen [0]{}%
\providecommand \bibitemStop [0]{}%
\providecommand \bibitemNoStop [0]{.\EOS\space}%
\providecommand \EOS [0]{\spacefactor3000\relax}%
\providecommand \BibitemShut  [1]{\csname bibitem#1\endcsname}%
\let\auto@bib@innerbib\@empty
%</preamble>
\bibitem [{\citenamefont {{\v{Z}}uti{\'c}}\ \emph {et~al.}(2004)\citenamefont
  {{\v{Z}}uti{\'c}}, \citenamefont {Fabian},\ and\ \citenamefont
  {Sarma}}]{vzutic2004spintronics}%
  \BibitemOpen
  \bibfield  {author} {\bibinfo {author} {\bibfnamefont {I.}~\bibnamefont
  {{\v{Z}}uti{\'c}}}, \bibinfo {author} {\bibfnamefont {J.}~\bibnamefont
  {Fabian}},\ and\ \bibinfo {author} {\bibfnamefont {S.~D.}\ \bibnamefont
  {Sarma}},\ }\bibfield  {title} {\bibinfo {title} {Spintronics: Fundamentals
  and applications},\ }\href@noop {} {\bibfield  {journal} {\bibinfo  {journal}
  {Reviews of modern physics}\ }\textbf {\bibinfo {volume} {76}},\ \bibinfo
  {pages} {323} (\bibinfo {year} {2004})}\BibitemShut {NoStop}%
\bibitem [{\citenamefont {Nagaosa}\ \emph {et~al.}(2010)\citenamefont
  {Nagaosa}, \citenamefont {Sinova}, \citenamefont {Onoda}, \citenamefont
  {MacDonald},\ and\ \citenamefont {Ong}}]{nagaosa2010anomalous}%
  \BibitemOpen
  \bibfield  {author} {\bibinfo {author} {\bibfnamefont {N.}~\bibnamefont
  {Nagaosa}}, \bibinfo {author} {\bibfnamefont {J.}~\bibnamefont {Sinova}},
  \bibinfo {author} {\bibfnamefont {S.}~\bibnamefont {Onoda}}, \bibinfo
  {author} {\bibfnamefont {A.~H.}\ \bibnamefont {MacDonald}},\ and\ \bibinfo
  {author} {\bibfnamefont {N.~P.}\ \bibnamefont {Ong}},\ }\bibfield  {title}
  {\bibinfo {title} {Anomalous hall effect},\ }\href@noop {} {\bibfield
  {journal} {\bibinfo  {journal} {Reviews of modern physics}\ }\textbf
  {\bibinfo {volume} {82}},\ \bibinfo {pages} {1539} (\bibinfo {year}
  {2010})}\BibitemShut {NoStop}%
\bibitem [{\citenamefont {Behnia}\ and\ \citenamefont
  {Aubin}(2016)}]{behnia2016nernst}%
  \BibitemOpen
  \bibfield  {author} {\bibinfo {author} {\bibfnamefont {K.}~\bibnamefont
  {Behnia}}\ and\ \bibinfo {author} {\bibfnamefont {H.}~\bibnamefont {Aubin}},\
  }\bibfield  {title} {\bibinfo {title} {Nernst effect in metals and
  superconductors: a review of concepts and experiments},\ }\href@noop {}
  {\bibfield  {journal} {\bibinfo  {journal} {Reports on Progress in Physics}\
  }\textbf {\bibinfo {volume} {79}},\ \bibinfo {pages} {046502} (\bibinfo
  {year} {2016})}\BibitemShut {NoStop}%
\bibitem [{\citenamefont {Lee}\ \emph {et~al.}(2004)\citenamefont {Lee},
  \citenamefont {Watauchi}, \citenamefont {Miller}, \citenamefont {Cava},\ and\
  \citenamefont {Ong}}]{lee2004anomalous}%
  \BibitemOpen
  \bibfield  {author} {\bibinfo {author} {\bibfnamefont {W.-L.}\ \bibnamefont
  {Lee}}, \bibinfo {author} {\bibfnamefont {S.}~\bibnamefont {Watauchi}},
  \bibinfo {author} {\bibfnamefont {V.}~\bibnamefont {Miller}}, \bibinfo
  {author} {\bibfnamefont {R.}~\bibnamefont {Cava}},\ and\ \bibinfo {author}
  {\bibfnamefont {N.}~\bibnamefont {Ong}},\ }\bibfield  {title} {\bibinfo
  {title} {Anomalous hall heat current and nernst effect in the c u c r 2 s e
  4- x b r x ferromagnet},\ }\href@noop {} {\bibfield  {journal} {\bibinfo
  {journal} {Physical review letters}\ }\textbf {\bibinfo {volume} {93}},\
  \bibinfo {pages} {226601} (\bibinfo {year} {2004})}\BibitemShut {NoStop}%
\bibitem [{\citenamefont {Chen}\ \emph {et~al.}(2014)\citenamefont {Chen},
  \citenamefont {Niu},\ and\ \citenamefont {MacDonald}}]{chen2014anomalous}%
  \BibitemOpen
  \bibfield  {author} {\bibinfo {author} {\bibfnamefont {H.}~\bibnamefont
  {Chen}}, \bibinfo {author} {\bibfnamefont {Q.}~\bibnamefont {Niu}},\ and\
  \bibinfo {author} {\bibfnamefont {A.~H.}\ \bibnamefont {MacDonald}},\
  }\bibfield  {title} {\bibinfo {title} {Anomalous hall effect arising from
  noncollinear antiferromagnetism},\ }\href@noop {} {\bibfield  {journal}
  {\bibinfo  {journal} {Physical review letters}\ }\textbf {\bibinfo {volume}
  {112}},\ \bibinfo {pages} {017205} (\bibinfo {year} {2014})}\BibitemShut
  {NoStop}%
\bibitem [{\citenamefont {Zhang}\ \emph
  {et~al.}(2018{\natexlab{a}})\citenamefont {Zhang}, \citenamefont
  {{\v{Z}}elezn{\`y}}, \citenamefont {Sun}, \citenamefont {Van Den~Brink},\
  and\ \citenamefont {Yan}}]{zhang2018spin}%
  \BibitemOpen
  \bibfield  {author} {\bibinfo {author} {\bibfnamefont {Y.}~\bibnamefont
  {Zhang}}, \bibinfo {author} {\bibfnamefont {J.}~\bibnamefont
  {{\v{Z}}elezn{\`y}}}, \bibinfo {author} {\bibfnamefont {Y.}~\bibnamefont
  {Sun}}, \bibinfo {author} {\bibfnamefont {J.}~\bibnamefont {Van Den~Brink}},\
  and\ \bibinfo {author} {\bibfnamefont {B.}~\bibnamefont {Yan}},\ }\bibfield
  {title} {\bibinfo {title} {Spin hall effect emerging from a noncollinear
  magnetic lattice without spin--orbit coupling},\ }\href@noop {} {\bibfield
  {journal} {\bibinfo  {journal} {New Journal of Physics}\ }\textbf {\bibinfo
  {volume} {20}},\ \bibinfo {pages} {073028} (\bibinfo {year}
  {2018}{\natexlab{a}})}\BibitemShut {NoStop}%
\bibitem [{\citenamefont {Nayak}\ \emph {et~al.}(2016)\citenamefont {Nayak},
  \citenamefont {Fischer}, \citenamefont {Sun}, \citenamefont {Yan},
  \citenamefont {Karel}, \citenamefont {Komarek}, \citenamefont {Shekhar},
  \citenamefont {Kumar}, \citenamefont {Schnelle}, \citenamefont {K{\"u}bler}
  \emph {et~al.}}]{nayak2016large}%
  \BibitemOpen
  \bibfield  {author} {\bibinfo {author} {\bibfnamefont {A.~K.}\ \bibnamefont
  {Nayak}}, \bibinfo {author} {\bibfnamefont {J.~E.}\ \bibnamefont {Fischer}},
  \bibinfo {author} {\bibfnamefont {Y.}~\bibnamefont {Sun}}, \bibinfo {author}
  {\bibfnamefont {B.}~\bibnamefont {Yan}}, \bibinfo {author} {\bibfnamefont
  {J.}~\bibnamefont {Karel}}, \bibinfo {author} {\bibfnamefont {A.~C.}\
  \bibnamefont {Komarek}}, \bibinfo {author} {\bibfnamefont {C.}~\bibnamefont
  {Shekhar}}, \bibinfo {author} {\bibfnamefont {N.}~\bibnamefont {Kumar}},
  \bibinfo {author} {\bibfnamefont {W.}~\bibnamefont {Schnelle}}, \bibinfo
  {author} {\bibfnamefont {J.}~\bibnamefont {K{\"u}bler}}, \emph {et~al.},\
  }\bibfield  {title} {\bibinfo {title} {Large anomalous hall effect driven by
  a nonvanishing berry curvature in the noncolinear antiferromagnet mn3ge},\
  }\href@noop {} {\bibfield  {journal} {\bibinfo  {journal} {Science advances}\
  }\textbf {\bibinfo {volume} {2}},\ \bibinfo {pages} {e1501870} (\bibinfo
  {year} {2016})}\BibitemShut {NoStop}%
\bibitem [{\citenamefont {Ganguly}\ \emph {et~al.}(2011)\citenamefont
  {Ganguly}, \citenamefont {Costa}, \citenamefont {Klautau}, \citenamefont
  {Bergman}, \citenamefont {Sanyal}, \citenamefont {Mookerjee},\ and\
  \citenamefont {Eriksson}}]{ganguly2011augmented}%
  \BibitemOpen
  \bibfield  {author} {\bibinfo {author} {\bibfnamefont {S.}~\bibnamefont
  {Ganguly}}, \bibinfo {author} {\bibfnamefont {M.}~\bibnamefont {Costa}},
  \bibinfo {author} {\bibfnamefont {A.~B.}\ \bibnamefont {Klautau}}, \bibinfo
  {author} {\bibfnamefont {A.}~\bibnamefont {Bergman}}, \bibinfo {author}
  {\bibfnamefont {B.}~\bibnamefont {Sanyal}}, \bibinfo {author} {\bibfnamefont
  {A.}~\bibnamefont {Mookerjee}},\ and\ \bibinfo {author} {\bibfnamefont
  {O.}~\bibnamefont {Eriksson}},\ }\bibfield  {title} {\bibinfo {title}
  {Augmented space recursion formulation of the study of disordered alloys with
  noncollinear magnetism and spin-orbit coupling: Application to mnpt and mn 3
  rh},\ }\href@noop {} {\bibfield  {journal} {\bibinfo  {journal} {Physical
  Review B}\ }\textbf {\bibinfo {volume} {83}},\ \bibinfo {pages} {094407}
  (\bibinfo {year} {2011})}\BibitemShut {NoStop}%
\bibitem [{\citenamefont {Yang}\ \emph {et~al.}(2017)\citenamefont {Yang},
  \citenamefont {Sun}, \citenamefont {Zhang}, \citenamefont {Shi},
  \citenamefont {Parkin},\ and\ \citenamefont {Yan}}]{yang2017topological}%
  \BibitemOpen
  \bibfield  {author} {\bibinfo {author} {\bibfnamefont {H.}~\bibnamefont
  {Yang}}, \bibinfo {author} {\bibfnamefont {Y.}~\bibnamefont {Sun}}, \bibinfo
  {author} {\bibfnamefont {Y.}~\bibnamefont {Zhang}}, \bibinfo {author}
  {\bibfnamefont {W.-J.}\ \bibnamefont {Shi}}, \bibinfo {author} {\bibfnamefont
  {S.~S.}\ \bibnamefont {Parkin}},\ and\ \bibinfo {author} {\bibfnamefont
  {B.}~\bibnamefont {Yan}},\ }\bibfield  {title} {\bibinfo {title} {Topological
  weyl semimetals in the chiral antiferromagnetic materials mn3ge and mn3sn},\
  }\href@noop {} {\bibfield  {journal} {\bibinfo  {journal} {New Journal of
  Physics}\ }\textbf {\bibinfo {volume} {19}},\ \bibinfo {pages} {015008}
  (\bibinfo {year} {2017})}\BibitemShut {NoStop}%
\bibitem [{\citenamefont {K{\"u}bler}\ and\ \citenamefont
  {Felser}(2014)}]{kubler2014non}%
  \BibitemOpen
  \bibfield  {author} {\bibinfo {author} {\bibfnamefont {J.}~\bibnamefont
  {K{\"u}bler}}\ and\ \bibinfo {author} {\bibfnamefont {C.}~\bibnamefont
  {Felser}},\ }\bibfield  {title} {\bibinfo {title} {Non-collinear
  antiferromagnets and the anomalous hall effect},\ }\href@noop {} {\bibfield
  {journal} {\bibinfo  {journal} {EPL (Europhysics Letters)}\ }\textbf
  {\bibinfo {volume} {108}},\ \bibinfo {pages} {67001} (\bibinfo {year}
  {2014})}\BibitemShut {NoStop}%
\bibitem [{\citenamefont {Li}\ \emph {et~al.}(2017)\citenamefont {Li},
  \citenamefont {Xu}, \citenamefont {Ding}, \citenamefont {Wang}, \citenamefont
  {Shen}, \citenamefont {Lu}, \citenamefont {Zhu},\ and\ \citenamefont
  {Behnia}}]{li2017anomalous}%
  \BibitemOpen
  \bibfield  {author} {\bibinfo {author} {\bibfnamefont {X.}~\bibnamefont
  {Li}}, \bibinfo {author} {\bibfnamefont {L.}~\bibnamefont {Xu}}, \bibinfo
  {author} {\bibfnamefont {L.}~\bibnamefont {Ding}}, \bibinfo {author}
  {\bibfnamefont {J.}~\bibnamefont {Wang}}, \bibinfo {author} {\bibfnamefont
  {M.}~\bibnamefont {Shen}}, \bibinfo {author} {\bibfnamefont {X.}~\bibnamefont
  {Lu}}, \bibinfo {author} {\bibfnamefont {Z.}~\bibnamefont {Zhu}},\ and\
  \bibinfo {author} {\bibfnamefont {K.}~\bibnamefont {Behnia}},\ }\bibfield
  {title} {\bibinfo {title} {Anomalous nernst and righi-leduc effects in mn 3
  sn: Berry curvature and entropy flow},\ }\href@noop {} {\bibfield  {journal}
  {\bibinfo  {journal} {Physical review letters}\ }\textbf {\bibinfo {volume}
  {119}},\ \bibinfo {pages} {056601} (\bibinfo {year} {2017})}\BibitemShut
  {NoStop}%
\bibitem [{\citenamefont {Guo}\ and\ \citenamefont
  {Wang}(2017)}]{guo2017large}%
  \BibitemOpen
  \bibfield  {author} {\bibinfo {author} {\bibfnamefont {G.-Y.}\ \bibnamefont
  {Guo}}\ and\ \bibinfo {author} {\bibfnamefont {T.-C.}\ \bibnamefont {Wang}},\
  }\bibfield  {title} {\bibinfo {title} {Large anomalous nernst and spin nernst
  effects in the noncollinear antiferromagnets mn 3 x (x= sn, ge, ga)},\
  }\href@noop {} {\bibfield  {journal} {\bibinfo  {journal} {Physical Review
  B}\ }\textbf {\bibinfo {volume} {96}},\ \bibinfo {pages} {224415} (\bibinfo
  {year} {2017})}\BibitemShut {NoStop}%
\bibitem [{\citenamefont {Kiyohara}\ \emph {et~al.}(2016)\citenamefont
  {Kiyohara}, \citenamefont {Tomita},\ and\ \citenamefont
  {Nakatsuji}}]{kiyohara2016giant}%
  \BibitemOpen
  \bibfield  {author} {\bibinfo {author} {\bibfnamefont {N.}~\bibnamefont
  {Kiyohara}}, \bibinfo {author} {\bibfnamefont {T.}~\bibnamefont {Tomita}},\
  and\ \bibinfo {author} {\bibfnamefont {S.}~\bibnamefont {Nakatsuji}},\
  }\bibfield  {title} {\bibinfo {title} {Giant anomalous hall effect in the
  chiral antiferromagnet mn 3 ge},\ }\href@noop {} {\bibfield  {journal}
  {\bibinfo  {journal} {Physical Review Applied}\ }\textbf {\bibinfo {volume}
  {5}},\ \bibinfo {pages} {064009} (\bibinfo {year} {2016})}\BibitemShut
  {NoStop}%
\bibitem [{\citenamefont {Ikhlas}\ \emph {et~al.}(2017)\citenamefont {Ikhlas},
  \citenamefont {Tomita}, \citenamefont {Koretsune}, \citenamefont {Suzuki},
  \citenamefont {Nishio-Hamane}, \citenamefont {Arita}, \citenamefont {Otani},\
  and\ \citenamefont {Nakatsuji}}]{ikhlas2017large}%
  \BibitemOpen
  \bibfield  {author} {\bibinfo {author} {\bibfnamefont {M.}~\bibnamefont
  {Ikhlas}}, \bibinfo {author} {\bibfnamefont {T.}~\bibnamefont {Tomita}},
  \bibinfo {author} {\bibfnamefont {T.}~\bibnamefont {Koretsune}}, \bibinfo
  {author} {\bibfnamefont {M.-T.}\ \bibnamefont {Suzuki}}, \bibinfo {author}
  {\bibfnamefont {D.}~\bibnamefont {Nishio-Hamane}}, \bibinfo {author}
  {\bibfnamefont {R.}~\bibnamefont {Arita}}, \bibinfo {author} {\bibfnamefont
  {Y.}~\bibnamefont {Otani}},\ and\ \bibinfo {author} {\bibfnamefont
  {S.}~\bibnamefont {Nakatsuji}},\ }\bibfield  {title} {\bibinfo {title} {Large
  anomalous nernst effect at room temperature in a chiral antiferromagnet},\
  }\href@noop {} {\bibfield  {journal} {\bibinfo  {journal} {Nature Physics}\
  }\textbf {\bibinfo {volume} {13}},\ \bibinfo {pages} {1085} (\bibinfo {year}
  {2017})}\BibitemShut {NoStop}%
\bibitem [{\citenamefont {{\v{Z}}elezn{\`y}}\ \emph {et~al.}(2017)\citenamefont
  {{\v{Z}}elezn{\`y}}, \citenamefont {Zhang}, \citenamefont {Felser},\ and\
  \citenamefont {Yan}}]{vzelezny2017spin}%
  \BibitemOpen
  \bibfield  {author} {\bibinfo {author} {\bibfnamefont {J.}~\bibnamefont
  {{\v{Z}}elezn{\`y}}}, \bibinfo {author} {\bibfnamefont {Y.}~\bibnamefont
  {Zhang}}, \bibinfo {author} {\bibfnamefont {C.}~\bibnamefont {Felser}},\ and\
  \bibinfo {author} {\bibfnamefont {B.}~\bibnamefont {Yan}},\ }\bibfield
  {title} {\bibinfo {title} {Spin-polarized current in noncollinear
  antiferromagnets},\ }\href@noop {} {\bibfield  {journal} {\bibinfo  {journal}
  {Physical review letters}\ }\textbf {\bibinfo {volume} {119}},\ \bibinfo
  {pages} {187204} (\bibinfo {year} {2017})}\BibitemShut {NoStop}%
\bibitem [{\citenamefont {Zhou}\ \emph {et~al.}(2020)\citenamefont {Zhou},
  \citenamefont {Hanke}, \citenamefont {Feng}, \citenamefont {Bl{\"u}gel},
  \citenamefont {Mokrousov},\ and\ \citenamefont {Yao}}]{zhou2020giant}%
  \BibitemOpen
  \bibfield  {author} {\bibinfo {author} {\bibfnamefont {X.}~\bibnamefont
  {Zhou}}, \bibinfo {author} {\bibfnamefont {J.-P.}\ \bibnamefont {Hanke}},
  \bibinfo {author} {\bibfnamefont {W.}~\bibnamefont {Feng}}, \bibinfo {author}
  {\bibfnamefont {S.}~\bibnamefont {Bl{\"u}gel}}, \bibinfo {author}
  {\bibfnamefont {Y.}~\bibnamefont {Mokrousov}},\ and\ \bibinfo {author}
  {\bibfnamefont {Y.}~\bibnamefont {Yao}},\ }\bibfield  {title} {\bibinfo
  {title} {Giant anomalous nernst effect in noncollinear antiferromagnetic
  mn-based antiperovskite nitrides},\ }\href@noop {} {\bibfield  {journal}
  {\bibinfo  {journal} {Physical review materials}\ }\textbf {\bibinfo {volume}
  {4}},\ \bibinfo {pages} {024408} (\bibinfo {year} {2020})}\BibitemShut
  {NoStop}%
\bibitem [{\citenamefont {Gurung}\ \emph {et~al.}(2019)\citenamefont {Gurung},
  \citenamefont {Shao}, \citenamefont {Paudel},\ and\ \citenamefont
  {Tsymbal}}]{gurung2019anomalous}%
  \BibitemOpen
  \bibfield  {author} {\bibinfo {author} {\bibfnamefont {G.}~\bibnamefont
  {Gurung}}, \bibinfo {author} {\bibfnamefont {D.-F.}\ \bibnamefont {Shao}},
  \bibinfo {author} {\bibfnamefont {T.~R.}\ \bibnamefont {Paudel}},\ and\
  \bibinfo {author} {\bibfnamefont {E.~Y.}\ \bibnamefont {Tsymbal}},\
  }\bibfield  {title} {\bibinfo {title} {Anomalous hall conductivity of
  noncollinear magnetic antiperovskites},\ }\href@noop {} {\bibfield  {journal}
  {\bibinfo  {journal} {Physical Review Materials}\ }\textbf {\bibinfo {volume}
  {3}},\ \bibinfo {pages} {044409} (\bibinfo {year} {2019})}\BibitemShut
  {NoStop}%
\bibitem [{\citenamefont {Huyen}\ \emph {et~al.}(2019)\citenamefont {Huyen},
  \citenamefont {Suzuki}, \citenamefont {Yamauchi},\ and\ \citenamefont
  {Oguchi}}]{huyen2019topology}%
  \BibitemOpen
  \bibfield  {author} {\bibinfo {author} {\bibfnamefont {V.~T.~N.}\
  \bibnamefont {Huyen}}, \bibinfo {author} {\bibfnamefont {M.-T.}\ \bibnamefont
  {Suzuki}}, \bibinfo {author} {\bibfnamefont {K.}~\bibnamefont {Yamauchi}},\
  and\ \bibinfo {author} {\bibfnamefont {T.}~\bibnamefont {Oguchi}},\
  }\bibfield  {title} {\bibinfo {title} {Topology analysis for anomalous hall
  effect in the noncollinear antiferromagnetic states of mn 3 a n (a= ni, cu,
  zn, ga, ge, pd, in, sn, ir, pt)},\ }\href@noop {} {\bibfield  {journal}
  {\bibinfo  {journal} {Physical Review B}\ }\textbf {\bibinfo {volume}
  {100}},\ \bibinfo {pages} {094426} (\bibinfo {year} {2019})}\BibitemShut
  {NoStop}%
\bibitem [{\citenamefont {Feng}\ \emph {et~al.}(2020)\citenamefont {Feng},
  \citenamefont {Zhou}, \citenamefont {{\v{S}}mejkal}, \citenamefont {Wu},
  \citenamefont {Zhu}, \citenamefont {Guo}, \citenamefont
  {Gonz{\'a}lez-Hern{\'a}ndez}, \citenamefont {Wang}, \citenamefont {Yan},
  \citenamefont {Qin} \emph {et~al.}}]{feng2020observation}%
  \BibitemOpen
  \bibfield  {author} {\bibinfo {author} {\bibfnamefont {Z.}~\bibnamefont
  {Feng}}, \bibinfo {author} {\bibfnamefont {X.}~\bibnamefont {Zhou}}, \bibinfo
  {author} {\bibfnamefont {L.}~\bibnamefont {{\v{S}}mejkal}}, \bibinfo {author}
  {\bibfnamefont {L.}~\bibnamefont {Wu}}, \bibinfo {author} {\bibfnamefont
  {Z.}~\bibnamefont {Zhu}}, \bibinfo {author} {\bibfnamefont {H.}~\bibnamefont
  {Guo}}, \bibinfo {author} {\bibfnamefont {R.}~\bibnamefont
  {Gonz{\'a}lez-Hern{\'a}ndez}}, \bibinfo {author} {\bibfnamefont
  {X.}~\bibnamefont {Wang}}, \bibinfo {author} {\bibfnamefont {H.}~\bibnamefont
  {Yan}}, \bibinfo {author} {\bibfnamefont {P.}~\bibnamefont {Qin}}, \emph
  {et~al.},\ }\bibfield  {title} {\bibinfo {title} {Observation of the
  anomalous hall effect in a collinear antiferromagnet},\ }\href@noop {}
  {\bibfield  {journal} {\bibinfo  {journal} {arXiv preprint arXiv:2002.08712}\
  } (\bibinfo {year} {2020})}\BibitemShut {NoStop}%
\bibitem [{\citenamefont {Hasegawa}\ \emph {et~al.}(2015)\citenamefont
  {Hasegawa}, \citenamefont {Mizuguchi}, \citenamefont {Sakuraba},
  \citenamefont {Kamada}, \citenamefont {Kojima}, \citenamefont {Kubota},
  \citenamefont {Mizukami}, \citenamefont {Miyazaki},\ and\ \citenamefont
  {Takanashi}}]{hasegawa2015material}%
  \BibitemOpen
  \bibfield  {author} {\bibinfo {author} {\bibfnamefont {K.}~\bibnamefont
  {Hasegawa}}, \bibinfo {author} {\bibfnamefont {M.}~\bibnamefont {Mizuguchi}},
  \bibinfo {author} {\bibfnamefont {Y.}~\bibnamefont {Sakuraba}}, \bibinfo
  {author} {\bibfnamefont {T.}~\bibnamefont {Kamada}}, \bibinfo {author}
  {\bibfnamefont {T.}~\bibnamefont {Kojima}}, \bibinfo {author} {\bibfnamefont
  {T.}~\bibnamefont {Kubota}}, \bibinfo {author} {\bibfnamefont
  {S.}~\bibnamefont {Mizukami}}, \bibinfo {author} {\bibfnamefont
  {T.}~\bibnamefont {Miyazaki}},\ and\ \bibinfo {author} {\bibfnamefont
  {K.}~\bibnamefont {Takanashi}},\ }\bibfield  {title} {\bibinfo {title}
  {Material dependence of anomalous nernst effect in perpendicularly magnetized
  ordered-alloy thin films},\ }\href@noop {} {\bibfield  {journal} {\bibinfo
  {journal} {Applied Physics Letters}\ }\textbf {\bibinfo {volume} {106}},\
  \bibinfo {pages} {252405} (\bibinfo {year} {2015})}\BibitemShut {NoStop}%
\bibitem [{\citenamefont {Huang}\ \emph {et~al.}(2011)\citenamefont {Huang},
  \citenamefont {Wang}, \citenamefont {Lee}, \citenamefont {Kwo},\ and\
  \citenamefont {Chien}}]{huang2011intrinsic}%
  \BibitemOpen
  \bibfield  {author} {\bibinfo {author} {\bibfnamefont {S.}~\bibnamefont
  {Huang}}, \bibinfo {author} {\bibfnamefont {W.}~\bibnamefont {Wang}},
  \bibinfo {author} {\bibfnamefont {S.}~\bibnamefont {Lee}}, \bibinfo {author}
  {\bibfnamefont {J.}~\bibnamefont {Kwo}},\ and\ \bibinfo {author}
  {\bibfnamefont {C.}~\bibnamefont {Chien}},\ }\bibfield  {title} {\bibinfo
  {title} {Intrinsic spin-dependent thermal transport},\ }\href@noop {}
  {\bibfield  {journal} {\bibinfo  {journal} {Physical review letters}\
  }\textbf {\bibinfo {volume} {107}},\ \bibinfo {pages} {216604} (\bibinfo
  {year} {2011})}\BibitemShut {NoStop}%
\bibitem [{\citenamefont {Johnson}(1975)}]{johnson1975diffusionless}%
  \BibitemOpen
  \bibfield  {author} {\bibinfo {author} {\bibfnamefont {V.}~\bibnamefont
  {Johnson}},\ }\bibfield  {title} {\bibinfo {title} {Diffusionless
  orthorhombic to hexagonal transitions in ternary silicides and germanides},\
  }\href@noop {} {\bibfield  {journal} {\bibinfo  {journal} {Inorganic
  chemistry}\ }\textbf {\bibinfo {volume} {14}},\ \bibinfo {pages} {1117}
  (\bibinfo {year} {1975})}\BibitemShut {NoStop}%
\bibitem [{\citenamefont {Szytula}\ \emph {et~al.}(1981)\citenamefont
  {Szytula}, \citenamefont {Pedziwiatr}, \citenamefont {Tomkowicz},\ and\
  \citenamefont {Bazela}}]{szytula1981crystal}%
  \BibitemOpen
  \bibfield  {author} {\bibinfo {author} {\bibfnamefont {A.}~\bibnamefont
  {Szytula}}, \bibinfo {author} {\bibfnamefont {A.}~\bibnamefont {Pedziwiatr}},
  \bibinfo {author} {\bibfnamefont {Z.}~\bibnamefont {Tomkowicz}},\ and\
  \bibinfo {author} {\bibfnamefont {W.}~\bibnamefont {Bazela}},\ }\bibfield
  {title} {\bibinfo {title} {Crystal and magnetic structure of comnge, cofege,
  femnge and nifege},\ }\href@noop {} {\bibfield  {journal} {\bibinfo
  {journal} {Journal of Magnetism and Magnetic Materials}\ }\textbf {\bibinfo
  {volume} {25}},\ \bibinfo {pages} {176} (\bibinfo {year} {1981})}\BibitemShut
  {NoStop}%
\bibitem [{\citenamefont {Jeitschko}(1975)}]{jeitschko1975high}%
  \BibitemOpen
  \bibfield  {author} {\bibinfo {author} {\bibfnamefont {W.}~\bibnamefont
  {Jeitschko}},\ }\bibfield  {title} {\bibinfo {title} {A high-temperature
  x-ray study of the displacive phase transition in mncoge},\ }\href@noop {}
  {\bibfield  {journal} {\bibinfo  {journal} {Acta Crystallographica Section B:
  Structural Crystallography and Crystal Chemistry}\ }\textbf {\bibinfo
  {volume} {31}},\ \bibinfo {pages} {1187} (\bibinfo {year}
  {1975})}\BibitemShut {NoStop}%
\bibitem [{\citenamefont {Yu}\ \emph {et~al.}(2006)\citenamefont {Yu},
  \citenamefont {Liu}, \citenamefont {Liu}, \citenamefont {Chen}, \citenamefont
  {Cao}, \citenamefont {Wu}, \citenamefont {Zhang},\ and\ \citenamefont
  {Zhang}}]{yu2006large}%
  \BibitemOpen
  \bibfield  {author} {\bibinfo {author} {\bibfnamefont {S.}~\bibnamefont
  {Yu}}, \bibinfo {author} {\bibfnamefont {Z.}~\bibnamefont {Liu}}, \bibinfo
  {author} {\bibfnamefont {G.}~\bibnamefont {Liu}}, \bibinfo {author}
  {\bibfnamefont {J.}~\bibnamefont {Chen}}, \bibinfo {author} {\bibfnamefont
  {Z.}~\bibnamefont {Cao}}, \bibinfo {author} {\bibfnamefont {G.}~\bibnamefont
  {Wu}}, \bibinfo {author} {\bibfnamefont {B.}~\bibnamefont {Zhang}},\ and\
  \bibinfo {author} {\bibfnamefont {X.}~\bibnamefont {Zhang}},\ }\bibfield
  {title} {\bibinfo {title} {Large magnetoresistance in single-crystalline ni
  50 mn 50- x in x alloys (x= 14--16) upon martensitic transformation},\
  }\href@noop {} {\bibfield  {journal} {\bibinfo  {journal} {Applied Physics
  Letters}\ }\textbf {\bibinfo {volume} {89}},\ \bibinfo {pages} {162503}
  (\bibinfo {year} {2006})}\BibitemShut {NoStop}%
\bibitem [{\citenamefont {Barandiar{\'a}n}\ \emph {et~al.}(2009)\citenamefont
  {Barandiar{\'a}n}, \citenamefont {Chernenko}, \citenamefont {L{\'a}zpita},
  \citenamefont {Guti{\'e}rrez},\ and\ \citenamefont
  {Feuchtwanger}}]{barandiaran2009effect}%
  \BibitemOpen
  \bibfield  {author} {\bibinfo {author} {\bibfnamefont {J.}~\bibnamefont
  {Barandiar{\'a}n}}, \bibinfo {author} {\bibfnamefont {V.}~\bibnamefont
  {Chernenko}}, \bibinfo {author} {\bibfnamefont {P.}~\bibnamefont
  {L{\'a}zpita}}, \bibinfo {author} {\bibfnamefont {J.}~\bibnamefont
  {Guti{\'e}rrez}},\ and\ \bibinfo {author} {\bibfnamefont {J.}~\bibnamefont
  {Feuchtwanger}},\ }\bibfield  {title} {\bibinfo {title} {Effect of
  martensitic transformation and magnetic field on transport properties of
  ni-mn-ga and ni-fe-ga heusler alloys},\ }\href@noop {} {\bibfield  {journal}
  {\bibinfo  {journal} {Physical Review B}\ }\textbf {\bibinfo {volume} {80}},\
  \bibinfo {pages} {104404} (\bibinfo {year} {2009})}\BibitemShut {NoStop}%
\bibitem [{\citenamefont {Ullakko}\ \emph {et~al.}(1996)\citenamefont
  {Ullakko}, \citenamefont {Huang}, \citenamefont {Kantner}, \citenamefont
  {O’handley},\ and\ \citenamefont {Kokorin}}]{ullakko1996large}%
  \BibitemOpen
  \bibfield  {author} {\bibinfo {author} {\bibfnamefont {K.}~\bibnamefont
  {Ullakko}}, \bibinfo {author} {\bibfnamefont {J.}~\bibnamefont {Huang}},
  \bibinfo {author} {\bibfnamefont {C.}~\bibnamefont {Kantner}}, \bibinfo
  {author} {\bibfnamefont {R.}~\bibnamefont {O’handley}},\ and\ \bibinfo
  {author} {\bibfnamefont {V.}~\bibnamefont {Kokorin}},\ }\bibfield  {title}
  {\bibinfo {title} {Large magnetic-field-induced strains in ni2mnga single
  crystals},\ }\href@noop {} {\bibfield  {journal} {\bibinfo  {journal}
  {Applied Physics Letters}\ }\textbf {\bibinfo {volume} {69}},\ \bibinfo
  {pages} {1966} (\bibinfo {year} {1996})}\BibitemShut {NoStop}%
\bibitem [{\citenamefont {Wu}\ \emph {et~al.}(1999)\citenamefont {Wu},
  \citenamefont {Yu}, \citenamefont {Meng}, \citenamefont {Chen}, \citenamefont
  {Yang}, \citenamefont {Qi}, \citenamefont {Zhan}, \citenamefont {Wang},
  \citenamefont {Zheng},\ and\ \citenamefont {Zhao}}]{wu1999giant}%
  \BibitemOpen
  \bibfield  {author} {\bibinfo {author} {\bibfnamefont {G.}~\bibnamefont
  {Wu}}, \bibinfo {author} {\bibfnamefont {C.}~\bibnamefont {Yu}}, \bibinfo
  {author} {\bibfnamefont {L.}~\bibnamefont {Meng}}, \bibinfo {author}
  {\bibfnamefont {J.}~\bibnamefont {Chen}}, \bibinfo {author} {\bibfnamefont
  {F.}~\bibnamefont {Yang}}, \bibinfo {author} {\bibfnamefont {S.}~\bibnamefont
  {Qi}}, \bibinfo {author} {\bibfnamefont {W.}~\bibnamefont {Zhan}}, \bibinfo
  {author} {\bibfnamefont {Z.}~\bibnamefont {Wang}}, \bibinfo {author}
  {\bibfnamefont {Y.}~\bibnamefont {Zheng}},\ and\ \bibinfo {author}
  {\bibfnamefont {L.}~\bibnamefont {Zhao}},\ }\bibfield  {title} {\bibinfo
  {title} {Giant magnetic-field-induced strains in heusler alloy nimnga with
  modified composition},\ }\href@noop {} {\bibfield  {journal} {\bibinfo
  {journal} {Applied Physics Letters}\ }\textbf {\bibinfo {volume} {75}},\
  \bibinfo {pages} {2990} (\bibinfo {year} {1999})}\BibitemShut {NoStop}%
\bibitem [{\citenamefont {Kainuma}\ \emph {et~al.}(2006)\citenamefont
  {Kainuma}, \citenamefont {Imano}, \citenamefont {Ito}, \citenamefont {Sutou},
  \citenamefont {Morito}, \citenamefont {Okamoto}, \citenamefont {Kitakami},
  \citenamefont {Oikawa}, \citenamefont {Fujita}, \citenamefont {Kanomata}
  \emph {et~al.}}]{kainuma2006magnetic}%
  \BibitemOpen
  \bibfield  {author} {\bibinfo {author} {\bibfnamefont {R.}~\bibnamefont
  {Kainuma}}, \bibinfo {author} {\bibfnamefont {Y.}~\bibnamefont {Imano}},
  \bibinfo {author} {\bibfnamefont {W.}~\bibnamefont {Ito}}, \bibinfo {author}
  {\bibfnamefont {Y.}~\bibnamefont {Sutou}}, \bibinfo {author} {\bibfnamefont
  {H.}~\bibnamefont {Morito}}, \bibinfo {author} {\bibfnamefont
  {S.}~\bibnamefont {Okamoto}}, \bibinfo {author} {\bibfnamefont
  {O.}~\bibnamefont {Kitakami}}, \bibinfo {author} {\bibfnamefont
  {K.}~\bibnamefont {Oikawa}}, \bibinfo {author} {\bibfnamefont
  {A.}~\bibnamefont {Fujita}}, \bibinfo {author} {\bibfnamefont
  {T.}~\bibnamefont {Kanomata}}, \emph {et~al.},\ }\bibfield  {title} {\bibinfo
  {title} {Magnetic-field-induced shape recovery by reverse phase
  transformation},\ }\href@noop {} {\bibfield  {journal} {\bibinfo  {journal}
  {Nature}\ }\textbf {\bibinfo {volume} {439}},\ \bibinfo {pages} {957}
  (\bibinfo {year} {2006})}\BibitemShut {NoStop}%
\bibitem [{\citenamefont {Krenke}\ \emph {et~al.}(2005)\citenamefont {Krenke},
  \citenamefont {Duman}, \citenamefont {Acet}, \citenamefont {Wassermann},
  \citenamefont {Moya}, \citenamefont {Ma{\~n}osa},\ and\ \citenamefont
  {Planes}}]{krenke2005inverse}%
  \BibitemOpen
  \bibfield  {author} {\bibinfo {author} {\bibfnamefont {T.}~\bibnamefont
  {Krenke}}, \bibinfo {author} {\bibfnamefont {E.}~\bibnamefont {Duman}},
  \bibinfo {author} {\bibfnamefont {M.}~\bibnamefont {Acet}}, \bibinfo {author}
  {\bibfnamefont {E.~F.}\ \bibnamefont {Wassermann}}, \bibinfo {author}
  {\bibfnamefont {X.}~\bibnamefont {Moya}}, \bibinfo {author} {\bibfnamefont
  {L.}~\bibnamefont {Ma{\~n}osa}},\ and\ \bibinfo {author} {\bibfnamefont
  {A.}~\bibnamefont {Planes}},\ }\bibfield  {title} {\bibinfo {title} {Inverse
  magnetocaloric effect in ferromagnetic ni--mn--sn alloys},\ }\href@noop {}
  {\bibfield  {journal} {\bibinfo  {journal} {Nature materials}\ }\textbf
  {\bibinfo {volume} {4}},\ \bibinfo {pages} {450} (\bibinfo {year}
  {2005})}\BibitemShut {NoStop}%
\bibitem [{\citenamefont {Gutfleisch}\ \emph {et~al.}(2011)\citenamefont
  {Gutfleisch}, \citenamefont {Willard}, \citenamefont {Br{\"u}ck},
  \citenamefont {Chen}, \citenamefont {Sankar},\ and\ \citenamefont
  {Liu}}]{gutfleisch2011magnetic}%
  \BibitemOpen
  \bibfield  {author} {\bibinfo {author} {\bibfnamefont {O.}~\bibnamefont
  {Gutfleisch}}, \bibinfo {author} {\bibfnamefont {M.~A.}\ \bibnamefont
  {Willard}}, \bibinfo {author} {\bibfnamefont {E.}~\bibnamefont {Br{\"u}ck}},
  \bibinfo {author} {\bibfnamefont {C.~H.}\ \bibnamefont {Chen}}, \bibinfo
  {author} {\bibfnamefont {S.}~\bibnamefont {Sankar}},\ and\ \bibinfo {author}
  {\bibfnamefont {J.~P.}\ \bibnamefont {Liu}},\ }\bibfield  {title} {\bibinfo
  {title} {Magnetic materials and devices for the 21st century: stronger,
  lighter, and more energy efficient},\ }\href@noop {} {\bibfield  {journal}
  {\bibinfo  {journal} {Advanced materials}\ }\textbf {\bibinfo {volume}
  {23}},\ \bibinfo {pages} {821} (\bibinfo {year} {2011})}\BibitemShut
  {NoStop}%
\bibitem [{\citenamefont {Tegus}\ \emph {et~al.}(2002)\citenamefont {Tegus},
  \citenamefont {Br{\"u}ck}, \citenamefont {Buschow},\ and\ \citenamefont
  {De~Boer}}]{tegus2002transition}%
  \BibitemOpen
  \bibfield  {author} {\bibinfo {author} {\bibfnamefont {O.}~\bibnamefont
  {Tegus}}, \bibinfo {author} {\bibfnamefont {E.}~\bibnamefont {Br{\"u}ck}},
  \bibinfo {author} {\bibfnamefont {K.}~\bibnamefont {Buschow}},\ and\ \bibinfo
  {author} {\bibfnamefont {F.}~\bibnamefont {De~Boer}},\ }\bibfield  {title}
  {\bibinfo {title} {Transition-metal-based magnetic refrigerants for
  room-temperature applications},\ }\href@noop {} {\bibfield  {journal}
  {\bibinfo  {journal} {Nature}\ }\textbf {\bibinfo {volume} {415}},\ \bibinfo
  {pages} {150} (\bibinfo {year} {2002})}\BibitemShut {NoStop}%
\bibitem [{\citenamefont {Liu}\ \emph {et~al.}(2012)\citenamefont {Liu},
  \citenamefont {Gottschall}, \citenamefont {Skokov}, \citenamefont {Moore},\
  and\ \citenamefont {Gutfleisch}}]{liu2012giant}%
  \BibitemOpen
  \bibfield  {author} {\bibinfo {author} {\bibfnamefont {J.}~\bibnamefont
  {Liu}}, \bibinfo {author} {\bibfnamefont {T.}~\bibnamefont {Gottschall}},
  \bibinfo {author} {\bibfnamefont {K.~P.}\ \bibnamefont {Skokov}}, \bibinfo
  {author} {\bibfnamefont {J.~D.}\ \bibnamefont {Moore}},\ and\ \bibinfo
  {author} {\bibfnamefont {O.}~\bibnamefont {Gutfleisch}},\ }\bibfield  {title}
  {\bibinfo {title} {Giant magnetocaloric effect driven by structural
  transitions},\ }\href@noop {} {\bibfield  {journal} {\bibinfo  {journal}
  {Nature materials}\ }\textbf {\bibinfo {volume} {11}},\ \bibinfo {pages}
  {620} (\bibinfo {year} {2012})}\BibitemShut {NoStop}%
\bibitem [{\citenamefont {Liu}\ \emph {et~al.}(2013)\citenamefont {Liu},
  \citenamefont {Zhang}, \citenamefont {Xu}, \citenamefont {Zhang},
  \citenamefont {Ma}, \citenamefont {Wang}, \citenamefont {Chen}, \citenamefont
  {Zhang}, \citenamefont {Wu}, \citenamefont {Feng} \emph
  {et~al.}}]{liu2013giant}%
  \BibitemOpen
  \bibfield  {author} {\bibinfo {author} {\bibfnamefont {E.}~\bibnamefont
  {Liu}}, \bibinfo {author} {\bibfnamefont {H.}~\bibnamefont {Zhang}}, \bibinfo
  {author} {\bibfnamefont {G.}~\bibnamefont {Xu}}, \bibinfo {author}
  {\bibfnamefont {X.}~\bibnamefont {Zhang}}, \bibinfo {author} {\bibfnamefont
  {R.}~\bibnamefont {Ma}}, \bibinfo {author} {\bibfnamefont {W.}~\bibnamefont
  {Wang}}, \bibinfo {author} {\bibfnamefont {J.}~\bibnamefont {Chen}}, \bibinfo
  {author} {\bibfnamefont {H.}~\bibnamefont {Zhang}}, \bibinfo {author}
  {\bibfnamefont {G.}~\bibnamefont {Wu}}, \bibinfo {author} {\bibfnamefont
  {L.}~\bibnamefont {Feng}}, \emph {et~al.},\ }\bibfield  {title} {\bibinfo
  {title} {Giant magnetocaloric effect in isostructural mnnige-conige system by
  establishing a curie-temperature window},\ }\href@noop {} {\bibfield
  {journal} {\bibinfo  {journal} {Applied Physics Letters}\ }\textbf {\bibinfo
  {volume} {102}},\ \bibinfo {pages} {122405} (\bibinfo {year}
  {2013})}\BibitemShut {NoStop}%
\bibitem [{\citenamefont {Glanz}(1998)}]{glanz1998making}%
  \BibitemOpen
  \bibfield  {author} {\bibinfo {author} {\bibfnamefont {J.}~\bibnamefont
  {Glanz}},\ }\bibfield  {title} {\bibinfo {title} {Making a bigger chill with
  magnets},\ }\href@noop {} {\bibfield  {journal} {\bibinfo  {journal}
  {Science}\ }\textbf {\bibinfo {volume} {279}},\ \bibinfo {pages} {2045}
  (\bibinfo {year} {1998})}\BibitemShut {NoStop}%
\bibitem [{\citenamefont {GschneidnerJr}\ \emph {et~al.}(2005)\citenamefont
  {GschneidnerJr}, \citenamefont {Pecharsky},\ and\ \citenamefont
  {Tsokol}}]{gschneidnerjr2005recent}%
  \BibitemOpen
  \bibfield  {author} {\bibinfo {author} {\bibfnamefont {K.~A.}\ \bibnamefont
  {GschneidnerJr}}, \bibinfo {author} {\bibfnamefont {V.}~\bibnamefont
  {Pecharsky}},\ and\ \bibinfo {author} {\bibfnamefont {A.}~\bibnamefont
  {Tsokol}},\ }\bibfield  {title} {\bibinfo {title} {Recent developments in
  magnetocaloric materials},\ }\href@noop {} {\bibfield  {journal} {\bibinfo
  {journal} {Reports on progress in physics}\ }\textbf {\bibinfo {volume}
  {68}},\ \bibinfo {pages} {1479} (\bibinfo {year} {2005})}\BibitemShut
  {NoStop}%
\bibitem [{\citenamefont {Singh}\ \emph {et~al.}(2021)\citenamefont {Singh},
  \citenamefont {Noky}, \citenamefont {Bhattacharya}, \citenamefont {Vir},
  \citenamefont {Sun}, \citenamefont {Kumar}, \citenamefont {Felser},\ and\
  \citenamefont {Shekhar}}]{singh2021anisotropic}%
  \BibitemOpen
  \bibfield  {author} {\bibinfo {author} {\bibfnamefont {S.}~\bibnamefont
  {Singh}}, \bibinfo {author} {\bibfnamefont {J.}~\bibnamefont {Noky}},
  \bibinfo {author} {\bibfnamefont {S.}~\bibnamefont {Bhattacharya}}, \bibinfo
  {author} {\bibfnamefont {P.}~\bibnamefont {Vir}}, \bibinfo {author}
  {\bibfnamefont {Y.}~\bibnamefont {Sun}}, \bibinfo {author} {\bibfnamefont
  {N.}~\bibnamefont {Kumar}}, \bibinfo {author} {\bibfnamefont
  {C.}~\bibnamefont {Felser}},\ and\ \bibinfo {author} {\bibfnamefont
  {C.}~\bibnamefont {Shekhar}},\ }\bibfield  {title} {\bibinfo {title}
  {Anisotropic nodal-line-derived large anomalous hall conductivity in zrmnp
  and hfmnp},\ }\href@noop {} {\bibfield  {journal} {\bibinfo  {journal}
  {Advanced Materials}\ ,\ \bibinfo {pages} {2104126}} (\bibinfo {year}
  {2021})}\BibitemShut {NoStop}%
\bibitem [{\citenamefont {Lamichhane}\ \emph {et~al.}(2016)\citenamefont
  {Lamichhane}, \citenamefont {Taufour}, \citenamefont {Masters}, \citenamefont
  {Parker}, \citenamefont {Kaluarachchi}, \citenamefont {Thimmaiah},
  \citenamefont {Bud'ko},\ and\ \citenamefont
  {Canfield}}]{lamichhane2016discovery}%
  \BibitemOpen
  \bibfield  {author} {\bibinfo {author} {\bibfnamefont {T.~N.}\ \bibnamefont
  {Lamichhane}}, \bibinfo {author} {\bibfnamefont {V.}~\bibnamefont {Taufour}},
  \bibinfo {author} {\bibfnamefont {M.~W.}\ \bibnamefont {Masters}}, \bibinfo
  {author} {\bibfnamefont {D.~S.}\ \bibnamefont {Parker}}, \bibinfo {author}
  {\bibfnamefont {U.~S.}\ \bibnamefont {Kaluarachchi}}, \bibinfo {author}
  {\bibfnamefont {S.}~\bibnamefont {Thimmaiah}}, \bibinfo {author}
  {\bibfnamefont {S.~L.}\ \bibnamefont {Bud'ko}},\ and\ \bibinfo {author}
  {\bibfnamefont {P.~C.}\ \bibnamefont {Canfield}},\ }\bibfield  {title}
  {\bibinfo {title} {Discovery of ferromagnetism with large magnetic anisotropy
  in zrmnp and hfmnp},\ }\href@noop {} {\bibfield  {journal} {\bibinfo
  {journal} {Applied Physics Letters}\ }\textbf {\bibinfo {volume} {109}},\
  \bibinfo {pages} {092402} (\bibinfo {year} {2016})}\BibitemShut {NoStop}%
\bibitem [{\citenamefont {Kresse}\ and\ \citenamefont
  {Hafner}(1993)}]{Kresse:1993}%
  \BibitemOpen
  \bibfield  {author} {\bibinfo {author} {\bibfnamefont {G.}~\bibnamefont
  {Kresse}}\ and\ \bibinfo {author} {\bibfnamefont {J.}~\bibnamefont
  {Hafner}},\ }\bibfield  {title} {\bibinfo {title} {Ab initio molecular
  dynamics for liquid metals},\ }\href@noop {} {\bibfield  {journal} {\bibinfo
  {journal} {Physical Review B}\ }\textbf {\bibinfo {volume} {47}},\ \bibinfo
  {pages} {558} (\bibinfo {year} {1993})}\BibitemShut {NoStop}%
\bibitem [{\citenamefont {Perdew}\ \emph {et~al.}(1996)\citenamefont {Perdew},
  \citenamefont {Burke},\ and\ \citenamefont {Ernzerhof}}]{Perdew:1996}%
  \BibitemOpen
  \bibfield  {author} {\bibinfo {author} {\bibfnamefont {J.~P.}\ \bibnamefont
  {Perdew}}, \bibinfo {author} {\bibfnamefont {K.}~\bibnamefont {Burke}},\ and\
  \bibinfo {author} {\bibfnamefont {M.}~\bibnamefont {Ernzerhof}},\ }\bibfield
  {title} {\bibinfo {title} {Generalized gradient approximation made simple},\
  }\href@noop {} {\bibfield  {journal} {\bibinfo  {journal} {Physical review
  letters}\ }\textbf {\bibinfo {volume} {77}},\ \bibinfo {pages} {3865}
  (\bibinfo {year} {1996})}\BibitemShut {NoStop}%
\bibitem [{\citenamefont {Mostofi}\ \emph {et~al.}(2008)\citenamefont
  {Mostofi}, \citenamefont {Yates}, \citenamefont {Lee}, \citenamefont {Souza},
  \citenamefont {Vanderbilt},\ and\ \citenamefont {Marzari}}]{Mostofi:2008}%
  \BibitemOpen
  \bibfield  {author} {\bibinfo {author} {\bibfnamefont {A.~A.}\ \bibnamefont
  {Mostofi}}, \bibinfo {author} {\bibfnamefont {J.~R.}\ \bibnamefont {Yates}},
  \bibinfo {author} {\bibfnamefont {Y.-S.}\ \bibnamefont {Lee}}, \bibinfo
  {author} {\bibfnamefont {I.}~\bibnamefont {Souza}}, \bibinfo {author}
  {\bibfnamefont {D.}~\bibnamefont {Vanderbilt}},\ and\ \bibinfo {author}
  {\bibfnamefont {N.}~\bibnamefont {Marzari}},\ }\bibfield  {title} {\bibinfo
  {title} {wannier90: A tool for obtaining maximally-localised wannier
  functions},\ }\href
  {https://doi.org/https://doi.org/10.1016/j.cpc.2007.11.016} {\bibfield
  {journal} {\bibinfo  {journal} {Computer Physics Communications}\ }\textbf
  {\bibinfo {volume} {178}},\ \bibinfo {pages} {685 } (\bibinfo {year}
  {2008})}\BibitemShut {NoStop}%
\bibitem [{\citenamefont {Zhang}\ \emph
  {et~al.}(2018{\natexlab{b}})\citenamefont {Zhang}, \citenamefont {Zhang},
  \citenamefont {Li}, \citenamefont {Koepernik}, \citenamefont {Yao},\ and\
  \citenamefont {Zhang}}]{Zeying:2018}%
  \BibitemOpen
  \bibfield  {author} {\bibinfo {author} {\bibfnamefont {Z.}~\bibnamefont
  {Zhang}}, \bibinfo {author} {\bibfnamefont {R.-W.}\ \bibnamefont {Zhang}},
  \bibinfo {author} {\bibfnamefont {X.}~\bibnamefont {Li}}, \bibinfo {author}
  {\bibfnamefont {K.}~\bibnamefont {Koepernik}}, \bibinfo {author}
  {\bibfnamefont {Y.}~\bibnamefont {Yao}},\ and\ \bibinfo {author}
  {\bibfnamefont {H.}~\bibnamefont {Zhang}},\ }\bibfield  {title} {\bibinfo
  {title} {High-throughput screening and automated processing toward novel
  topological insulators},\ }\href@noop {} {\bibfield  {journal} {\bibinfo
  {journal} {The journal of physical chemistry letters}\ }\textbf {\bibinfo
  {volume} {9}},\ \bibinfo {pages} {6224} (\bibinfo {year}
  {2018}{\natexlab{b}})}\BibitemShut {NoStop}%
\bibitem [{\citenamefont {Wu}\ \emph {et~al.}(2018)\citenamefont {Wu},
  \citenamefont {Zhang}, \citenamefont {Song}, \citenamefont {Troyer},\ and\
  \citenamefont {Soluyanov}}]{wanniertools:2018}%
  \BibitemOpen
  \bibfield  {author} {\bibinfo {author} {\bibfnamefont {Q.}~\bibnamefont
  {Wu}}, \bibinfo {author} {\bibfnamefont {S.}~\bibnamefont {Zhang}}, \bibinfo
  {author} {\bibfnamefont {H.-F.}\ \bibnamefont {Song}}, \bibinfo {author}
  {\bibfnamefont {M.}~\bibnamefont {Troyer}},\ and\ \bibinfo {author}
  {\bibfnamefont {A.~A.}\ \bibnamefont {Soluyanov}},\ }\bibfield  {title}
  {\bibinfo {title} {Wanniertools: An open-source software package for novel
  topological materials},\ }\href@noop {} {\bibfield  {journal} {\bibinfo
  {journal} {Computer Physics Communications}\ }\textbf {\bibinfo {volume}
  {224}},\ \bibinfo {pages} {405} (\bibinfo {year} {2018})}\BibitemShut
  {NoStop}%
\bibitem [{\citenamefont {Xiao}\ \emph {et~al.}(2010)\citenamefont {Xiao},
  \citenamefont {Chang},\ and\ \citenamefont {Niu}}]{Xiao:2010}%
  \BibitemOpen
  \bibfield  {author} {\bibinfo {author} {\bibfnamefont {D.}~\bibnamefont
  {Xiao}}, \bibinfo {author} {\bibfnamefont {M.-C.}\ \bibnamefont {Chang}},\
  and\ \bibinfo {author} {\bibfnamefont {Q.}~\bibnamefont {Niu}},\ }\bibfield
  {title} {\bibinfo {title} {Berry phase effects on electronic properties},\
  }\href {https://doi.org/10.1103/RevModPhys.82.1959} {\bibfield  {journal}
  {\bibinfo  {journal} {Rev. Mod. Phys.}\ }\textbf {\bibinfo {volume} {82}},\
  \bibinfo {pages} {1959} (\bibinfo {year} {2010})}\BibitemShut {NoStop}%
\bibitem [{\citenamefont {Samathrakis}\ \emph {et~al.}(2021)\citenamefont
  {Samathrakis}, \citenamefont {Long}, \citenamefont {Zhang}, \citenamefont
  {Singh},\ and\ \citenamefont {Zhang}}]{samathrakis2021enhanced}%
  \BibitemOpen
  \bibfield  {author} {\bibinfo {author} {\bibfnamefont {I.}~\bibnamefont
  {Samathrakis}}, \bibinfo {author} {\bibfnamefont {T.}~\bibnamefont {Long}},
  \bibinfo {author} {\bibfnamefont {Z.}~\bibnamefont {Zhang}}, \bibinfo
  {author} {\bibfnamefont {H.~K.}\ \bibnamefont {Singh}},\ and\ \bibinfo
  {author} {\bibfnamefont {H.}~\bibnamefont {Zhang}},\ }\bibfield  {title}
  {\bibinfo {title} {Enhanced anomalous nernst effects in ferromagnetic
  materials driven by weyl nodes},\ }\href
  {https://doi.org/10.1088/1361-6463/ac3351} {\bibfield  {journal} {\bibinfo
  {journal} {Journal of Physics D: Applied Physics}\ }\textbf {\bibinfo
  {volume} {55}},\ \bibinfo {pages} {074003} (\bibinfo {year}
  {2021})}\BibitemShut {NoStop}%
\bibitem [{\citenamefont {Wang}\ \emph {et~al.}(2018)\citenamefont {Wang},
  \citenamefont {Xu}, \citenamefont {Lou}, \citenamefont {Liu}, \citenamefont
  {Li}, \citenamefont {Huang}, \citenamefont {Shen}, \citenamefont {Weng},
  \citenamefont {Wang},\ and\ \citenamefont {Lei}}]{wang2018large}%
  \BibitemOpen
  \bibfield  {author} {\bibinfo {author} {\bibfnamefont {Q.}~\bibnamefont
  {Wang}}, \bibinfo {author} {\bibfnamefont {Y.}~\bibnamefont {Xu}}, \bibinfo
  {author} {\bibfnamefont {R.}~\bibnamefont {Lou}}, \bibinfo {author}
  {\bibfnamefont {Z.}~\bibnamefont {Liu}}, \bibinfo {author} {\bibfnamefont
  {M.}~\bibnamefont {Li}}, \bibinfo {author} {\bibfnamefont {Y.}~\bibnamefont
  {Huang}}, \bibinfo {author} {\bibfnamefont {D.}~\bibnamefont {Shen}},
  \bibinfo {author} {\bibfnamefont {H.}~\bibnamefont {Weng}}, \bibinfo {author}
  {\bibfnamefont {S.}~\bibnamefont {Wang}},\ and\ \bibinfo {author}
  {\bibfnamefont {H.}~\bibnamefont {Lei}},\ }\bibfield  {title} {\bibinfo
  {title} {Large intrinsic anomalous hall effect in half-metallic ferromagnet
  co 3 sn 2 s 2 with magnetic weyl fermions},\ }\href@noop {} {\bibfield
  {journal} {\bibinfo  {journal} {Nature communications}\ }\textbf {\bibinfo
  {volume} {9}},\ \bibinfo {pages} {1} (\bibinfo {year} {2018})}\BibitemShut
  {NoStop}%
\bibitem [{\citenamefont {Noky}\ \emph {et~al.}(2020)\citenamefont {Noky},
  \citenamefont {Zhang}, \citenamefont {Gooth}, \citenamefont {Felser},\ and\
  \citenamefont {Sun}}]{noky2020giant}%
  \BibitemOpen
  \bibfield  {author} {\bibinfo {author} {\bibfnamefont {J.}~\bibnamefont
  {Noky}}, \bibinfo {author} {\bibfnamefont {Y.}~\bibnamefont {Zhang}},
  \bibinfo {author} {\bibfnamefont {J.}~\bibnamefont {Gooth}}, \bibinfo
  {author} {\bibfnamefont {C.}~\bibnamefont {Felser}},\ and\ \bibinfo {author}
  {\bibfnamefont {Y.}~\bibnamefont {Sun}},\ }\bibfield  {title} {\bibinfo
  {title} {Giant anomalous hall and nernst effect in magnetic cubic heusler
  compounds},\ }\href@noop {} {\bibfield  {journal} {\bibinfo  {journal} {npj
  Computational Materials}\ }\textbf {\bibinfo {volume} {6}},\ \bibinfo {pages}
  {1} (\bibinfo {year} {2020})}\BibitemShut {NoStop}%
\bibitem [{\citenamefont {Singh}\ \emph {et~al.}(2022)\citenamefont {Singh},
  \citenamefont {Samathrakis}, \citenamefont {Shen},\ and\ \citenamefont
  {Zhang}}]{singh2022giant}%
  \BibitemOpen
  \bibfield  {author} {\bibinfo {author} {\bibfnamefont {H.~K.}\ \bibnamefont
  {Singh}}, \bibinfo {author} {\bibfnamefont {I.}~\bibnamefont {Samathrakis}},
  \bibinfo {author} {\bibfnamefont {C.}~\bibnamefont {Shen}},\ and\ \bibinfo
  {author} {\bibfnamefont {H.}~\bibnamefont {Zhang}},\ }\bibfield  {title}
  {\bibinfo {title} {Giant anomalous hall and anomalous nernst conductivities
  in antiperovskites and their tunability via magnetic fields},\ }\href@noop {}
  {\bibfield  {journal} {\bibinfo  {journal} {Physical Review Materials}\
  }\textbf {\bibinfo {volume} {6}},\ \bibinfo {pages} {045402} (\bibinfo {year}
  {2022})}\BibitemShut {NoStop}%
\bibitem [{\citenamefont {Zhou}\ \emph {et~al.}(2021)\citenamefont {Zhou},
  \citenamefont {Yamamoto}, \citenamefont {Miura}, \citenamefont {Iguchi},
  \citenamefont {Miura}, \citenamefont {Uchida},\ and\ \citenamefont
  {Sakuraba}}]{zhou2021seebeck}%
  \BibitemOpen
  \bibfield  {author} {\bibinfo {author} {\bibfnamefont {W.}~\bibnamefont
  {Zhou}}, \bibinfo {author} {\bibfnamefont {K.}~\bibnamefont {Yamamoto}},
  \bibinfo {author} {\bibfnamefont {A.}~\bibnamefont {Miura}}, \bibinfo
  {author} {\bibfnamefont {R.}~\bibnamefont {Iguchi}}, \bibinfo {author}
  {\bibfnamefont {Y.}~\bibnamefont {Miura}}, \bibinfo {author} {\bibfnamefont
  {K.-i.}\ \bibnamefont {Uchida}},\ and\ \bibinfo {author} {\bibfnamefont
  {Y.}~\bibnamefont {Sakuraba}},\ }\bibfield  {title} {\bibinfo {title}
  {Seebeck-driven transverse thermoelectric generation},\ }\href@noop {}
  {\bibfield  {journal} {\bibinfo  {journal} {Nature Materials}\ }\textbf
  {\bibinfo {volume} {20}},\ \bibinfo {pages} {463} (\bibinfo {year}
  {2021})}\BibitemShut {NoStop}%
\bibitem [{\citenamefont {Uchida}\ \emph {et~al.}(2021)\citenamefont {Uchida},
  \citenamefont {Zhou},\ and\ \citenamefont {Sakuraba}}]{uchida2021transverse}%
  \BibitemOpen
  \bibfield  {author} {\bibinfo {author} {\bibfnamefont {K.-i.}\ \bibnamefont
  {Uchida}}, \bibinfo {author} {\bibfnamefont {W.}~\bibnamefont {Zhou}},\ and\
  \bibinfo {author} {\bibfnamefont {Y.}~\bibnamefont {Sakuraba}},\ }\bibfield
  {title} {\bibinfo {title} {Transverse thermoelectric generation using
  magnetic materials},\ }\href@noop {} {\bibfield  {journal} {\bibinfo
  {journal} {Applied Physics Letters}\ }\textbf {\bibinfo {volume} {118}},\
  \bibinfo {pages} {140504} (\bibinfo {year} {2021})}\BibitemShut {NoStop}%
\bibitem [{\citenamefont {Sakai}\ \emph {et~al.}(2020)\citenamefont {Sakai},
  \citenamefont {Minami}, \citenamefont {Koretsune}, \citenamefont {Chen},
  \citenamefont {Higo}, \citenamefont {Wang}, \citenamefont {Nomoto},
  \citenamefont {Hirayama}, \citenamefont {Miwa}, \citenamefont {Nishio-Hamane}
  \emph {et~al.}}]{sakai2020iron}%
  \BibitemOpen
  \bibfield  {author} {\bibinfo {author} {\bibfnamefont {A.}~\bibnamefont
  {Sakai}}, \bibinfo {author} {\bibfnamefont {S.}~\bibnamefont {Minami}},
  \bibinfo {author} {\bibfnamefont {T.}~\bibnamefont {Koretsune}}, \bibinfo
  {author} {\bibfnamefont {T.}~\bibnamefont {Chen}}, \bibinfo {author}
  {\bibfnamefont {T.}~\bibnamefont {Higo}}, \bibinfo {author} {\bibfnamefont
  {Y.}~\bibnamefont {Wang}}, \bibinfo {author} {\bibfnamefont {T.}~\bibnamefont
  {Nomoto}}, \bibinfo {author} {\bibfnamefont {M.}~\bibnamefont {Hirayama}},
  \bibinfo {author} {\bibfnamefont {S.}~\bibnamefont {Miwa}}, \bibinfo {author}
  {\bibfnamefont {D.}~\bibnamefont {Nishio-Hamane}}, \emph {et~al.},\
  }\bibfield  {title} {\bibinfo {title} {Iron-based binary ferromagnets for
  transverse thermoelectric conversion},\ }\href@noop {} {\bibfield  {journal}
  {\bibinfo  {journal} {Nature}\ }\textbf {\bibinfo {volume} {581}},\ \bibinfo
  {pages} {53} (\bibinfo {year} {2020})}\BibitemShut {NoStop}%
\bibitem [{\citenamefont {Holanda}\ \emph {et~al.}(2021)\citenamefont
  {Holanda}, \citenamefont {Santos},\ and\ \citenamefont
  {Rezende}}]{holanda2021thermal}%
  \BibitemOpen
  \bibfield  {author} {\bibinfo {author} {\bibfnamefont {J.}~\bibnamefont
  {Holanda}}, \bibinfo {author} {\bibfnamefont {O.~A.}\ \bibnamefont
  {Santos}},\ and\ \bibinfo {author} {\bibfnamefont {S.~M.}\ \bibnamefont
  {Rezende}},\ }\bibfield  {title} {\bibinfo {title} {Thermal control of the
  intrinsic magnetic damping in a ferromagnetic metal},\ }\href@noop {}
  {\bibfield  {journal} {\bibinfo  {journal} {Physical Review Applied}\
  }\textbf {\bibinfo {volume} {16}},\ \bibinfo {pages} {014051} (\bibinfo
  {year} {2021})}\BibitemShut {NoStop}%
\bibitem [{\citenamefont {Seemann}\ \emph {et~al.}(2015)\citenamefont
  {Seemann}, \citenamefont {K{\"o}dderitzsch}, \citenamefont {Wimmer},\ and\
  \citenamefont {Ebert}}]{seemann2015symmetry}%
  \BibitemOpen
  \bibfield  {author} {\bibinfo {author} {\bibfnamefont {M.}~\bibnamefont
  {Seemann}}, \bibinfo {author} {\bibfnamefont {D.}~\bibnamefont
  {K{\"o}dderitzsch}}, \bibinfo {author} {\bibfnamefont {S.}~\bibnamefont
  {Wimmer}},\ and\ \bibinfo {author} {\bibfnamefont {H.}~\bibnamefont
  {Ebert}},\ }\bibfield  {title} {\bibinfo {title} {Symmetry-imposed shape of
  linear response tensors},\ }\href@noop {} {\bibfield  {journal} {\bibinfo
  {journal} {Physical Review B}\ }\textbf {\bibinfo {volume} {92}},\ \bibinfo
  {pages} {155138} (\bibinfo {year} {2015})}\BibitemShut {NoStop}%
\end{thebibliography}

%

\end{document}